# Ultima Thule: a Prediction for the Origin, Bulk Chemical Composition, and Physical Structure, submitted prior to the New Horizons Spacecraft 100 Pixel LORRI Data Return

Andrew J. R. Prentice [1]

[1] School of Physics & Astronomy, Monash University, Clayton VIC 3800, Australia

Email: Andrew.prentice@monash.edu

**Abstract**

The 2019 January 01 flypast of the Kuiper Belt object 2014 $MU_{69}$ (aka Ultima Thule) by the New Horizons spacecraft has provided the author with an important opportunity to test his 'gas ring' model of planetary origin, Prentice (1978). This model proposes that Ultima Thule and all Kuiper Belt objects, including proto-Pluto and Quaoar, condensed from the outermost gas ring that was shed by the contracting protosolar cloud, some 4560 Myr ago. A detailed account of the many unpublished improvements in the quantification of the gas ring model that have been made since 1978 are presented in the first half of this paper.

In the second half of the paper, I use the fully quantified gas ring model to compute the thermal properties of the gas ring in which Ultima condensed, and thence to predict the initial bulk chemical composition of the condensate. It is predicted that the KBOs initially contained large stores of $CO_2$ ice and $CH_4$ ices. These make up fractions 0.2210 and 0.0513 of the condensate mass, respectively. Water ice makes up a mass fraction 0.1845, nearly-dry rock has fraction 0.5269 and graphite has 0.0163. Next, I compute the thermal evolution of Ultima, taking into account the radiogenic heat released by the decay of $^{26}Al$. Stellar occultation data suggest that Ultima Thule may a bi-lobate object, consisting of 2 lobes of radius about 10 km and 7.5 km. The thermal evolution model shows that within 0.2 Myr, the peak internal temperatures are sufficient for a fraction ~0.7 of the $CH_4$ ice in the larger lobe to melt and for a fraction ~0.4 of the $CO_2$ ice to sublime. For the smaller lobe, these fractions are less. I propose that liquid $CH_4$ quickly migrates upwards to the surface and refreezes to form a thick and impervious outer shell of $CH_4$ ice. The sublimation of $CO_2$ ice takes place after the melting of $CH_4$. The possibility now exists for rising $CO_2$ vapour to become trapped beneath the $CH_4$ shell. This may lead to explosive eruptions of the outer crust and destruction of the primordial cratered surface of Ultima and loss of the $CO_2$. If 60% of $CO_2$ is lost, the lobe fractional radii each shrink by ~5%. Even so, the intensity of $^{26}Al$ radiogenic heating may not be sufficient to render the surface of Ultima Thule globally smooth unless the lobe sizes are of order ~15 km.

## 1 INTRODUCTION

The 2019 January 01 encounter of the New Horizons spacecraft with 2014 MU69 (unofficially named Ultima Thule) will provide the first ever closeup view of one of the most ancient and distant objects left over from the formation of our solar system. Ultima Thule, hereafter dubbed Ultima, is part of the so called Kuiper Belt of icy objects. It has a nearly-circular low-inclination orbit, with mean heliocentric distance of 44.2 AU, Porter et al. (2018). These orbital elements are very similar to those of 50000 Quaoar, which was one of the first and largest KBOs to be discovered, Trujillo & Brown (2003). Both Ultima and Quaoar are classified as cold classical KBOs (hereafter CCKBOs). The orbital radii of the CCKBOs cluster close to the radial mid-point of the Kuiper Belt, namely ~ 44 AU. The radial extent of the KB, namely 38 - 50 AU, was first predicted by Kuiper (1951) and later confirmed theoretically by Fernández (1980).

Ultima Thule is the second outer solar system body to be visited by the New Horizons spacecraft, following the spectacular flypast of the Pluto system in July 2015, Stern et al. (2015). Moore et al. (2018) have noted that the choice of Ultima Thule as the Kuiper Belt flyby object is most fortuitous as its size, approximately $20 \times 35$ km, is intermediate between that of comet nuclei, e.g. comet Hartley 2 whose 'double-lobe' dimensions are $0.7 \times 2.3$ km, A'Hearn et al. (2011), and Saturn's moon Phoebe whose mean diameter is $213 \pm 1.4$ km, Thomas et al. (2010). It is most likely that Phoebe was captured from the outer Solar System (hereafter OSS), Johnson & Lunine (2005). Pluto, of course is the largest of OSS bodies known to lie within 50 AU of the Sun. Its diameter is $2374 \pm 8$ km , Stern et al. (2015).

Very little is known physically about Ultima other than its approximate dimensions. Its oblong shape, as deduced from stellar occultation studies, suggests that it is a double-lobed body, Buie et al. (2017). Unpublished HST observations (see Moore et al. 2018) show that it is reddish in colour.

This implies the presence of organic compounds on the surface and hence a source of native hydrocarbons, most likely $CH_4$. Two of Pluto's moons Charon and Nix have red patches that may have formed by radiation processing of $CH_4$ ice, Grundy et al. (2016), Weaver et al. (2016).

The purpose of this pre-encounter paper is to present a model for the bulk chemical composition and physical structure of Ultima Thule. The model is based on the author's gas ring model for our Solar system's origin, Prentice (1978), (1984), (1993a), (2001), (2006), (2008), (2015), (2016), (2018). This model, which is also known as the Modern Laplacian Theory of Solar system origin – hereafter MLT – is described in more detail in section 2 below. It is a modern quantification of the original nebula hypothesis due to Laplace (1796). The basic premise of the MLT is that our planetary system condensed from a concentric family of orbiting gas rings. These rings were shed from the equator of the primordial rotating solar cloud, as a means for the cloud to dispose of excess spin angular momentum during its gravitational contraction. The initial radial size $R_e(0)$ of the cloud, which I dub the protosolar cloud (hereafter PSC), is assumed to exceed the present dimension of the solar system and to be of order 10,000 $R_\odot$, Prentice (1974), (1978). Here $R_\odot$ = 6.9598 × $10^{10}$ cm is the solar radius adopted in this paper. The present mean orbital distance of Ultima Thule is 9,500 $R_\odot$ (i.e. 44.2 AU) is thus consistent with its formation from the PSC. We shall see later on (Section 4.1) that Ultima and all of the CCKBOs initially condensed some ~ 18% closer to the Sun than where they are today. This is because of a secular expansion of the orbits arising from the loss of mass of the PSC during gravitational contraction.

In this paper I propose that all of the CCKBOs, especially Ultima and Quaoar, as well as proto-Pluto, condensed from the first and outermost gas ring that was shed by the PSC. Elsewhere I have suggested that the Pluto-Charon binary system was formed by the rotational fission of proto-Pluto, Prentice (1993b), (2015), (2016). All of these bodies should have thus started out having the same bulk chemical composition.

## 1.1 Layout of the paper

A brief discussion is given in Section 2 of the theoretical difficulties that face a modern quantified reformulation of the original Laplace hypothesis. This includes the need to understand how a single large rotating cloud, dubbed the Protosolar Cloud (PSC), can safely contract to present solar size without losing most of its mass. The importance of magnetic fields in ridding excess spin angular momentum of the PSC in the final stages of contraction, for equatorial sizes less than $15 R_\odot$, is also discussed. Section 3 provides a brief description of supersonic turbulent convection. This is the mechanism that helps circumvent the objections that led to the downfall of the original Laplacian nebula hypothesis. The 3 important parameters $\beta_0$, $F_0$ and $\theta_s$ that define the state of the PSC are introduced in Section 3. In Section 4, the procedure for solving for $\beta_0$, $F_0$ and $\theta_s$ is explained. The specification of these parameters is essential in order to quantify the gravitational contraction and thermal evolution of the PSC. The procedure for numerically modelling the radial contraction of the cloud as function of time $t$, starting from an initial equatorial size $R_e(0) = 10,000 R_\odot$, is established in Section 4. In Section 5, the computed values for the thermophysical state of the family of gas rings that are shed by the contracting PSC are presented. This includes the determination of the relative proportions of the 4 possible states of carbon, namely CO, $CO_2$, $CH_4$ and graphite C(s), that form in each gas ring. Of special interest are the properties of the outermost gas ring shed by the PSC and the prediction for the bulk chemical composition of the condensate at that orbit. Section 6 is devoted to computing the thermal evolution and physical state of Ultima Thule, starting with the bulk chemical compositions derived in Section 5. A previous computer program EVOL9CTC.FOR that was constructed by the author for studying the thermal evolution of the satellites of Jupiter and Saturn, as well as of Ceres, is extended to include the radiogenic heat that derives from the decay of the short-lived isotope [26]Al. The model is applied to Ultima and the two lobes that make up its expected bi-lobate structure. Predictions for internal structure, composition and surface appearance of Ultima are summarized in Section 7.

## 2 THE MODERN LAPLACIAN THEORY

### 2.1 Early problems with the Laplace hypothesis

The Laplace (1796) nebula hypothesis for the formation of the solar system is conceptually simple and attractive. It is not surprising, therefore, that it became widely accepted for almost a full century. The first serious criticism of the hypothesis was made by Babinet (1861). He noted that the mass of the planetary system is very small compared to that of the present-day Sun. If, as Laplace proposed, the primordial solar cloud had first become fully rotating at an equatorial radius $R_e = R_0$ that matched the outer edge of the planetary system, say for $R_0 = 10^4 R_\odot$, then in order for the cloud to safely contract to its present size $R_\odot$, it would have to give up a very large fraction of its mass, in addition to losing nearly all of its spin angular momentum. To appreciate this point consider the equations for the rotational evolution of the mass $M(R_e)$ and spin angular momentum $L(R_e)$ of a uniformly rotating cloud which maintains a Kepler angular velocity $\omega_{\text{Kep}}(R_e) = (GM_e/R_e^3)^{1/2}$ at its equator during contraction, Schatzman (1949). If $M_0$ and $L_0$ denote the initial mass and spin angular momentum of the cloud, we have

$M(R_e) = M_0 \cdot (R_e/R_0)^{\alpha f_e}$ and $L(R_e) = L_0 \cdot (R_e/R_0)^\alpha$ where $\alpha = 1/(2-3f_e)$ and $f_e$ is the polar axial moment-of-inertia factor. Taking $f_e = 0.135$, corresponding to a fully-rotating convective cloud of polytropic index $n = 1.5$, Auer & Woolf 1965, and final cloud radius $R_{\text{fin}} = 15 R_\odot$, we find $M_{\text{fin}} = 0.577 M_0$. That is, the cloud would have to give up almost half of its mass to contract to a size that is still 10 times larger than that of a fully-rotating solar globe having polar radius $R_\odot$.

Jeans (1929) was the first cosmogonist to point out that the mass-angular momentum difficulty of the Laplace hypothesis can be overcome if the protosolar cloud was very centrally condensed. This means that $f_e \ll 1$. Choosing $f_e = 0.02$, we get $M_{\text{fin}} = 0.935 M_0$ and hence that the mass of the nebula shed by the cloud – and from which the planetary system condensed – was $m_{\text{neb}} = 0.069 M_\odot$, setting $M_{\text{fin}} = M_\odot$. This value coincides with the lower bound estimate of $0.07 M_\odot$ derived by Weidenschilling (1977) in his cosmochemical reconstruction of the planetary nebular mass. It is also in line with the mass estimate $m_{\text{neb}} \simeq 0.10 M_\odot$ made by Kuiper (1951).

The real problem faced by any attempt to re-establish the Laplace hypothesis is (i) to find a mechanism for the PSC to become very centrally condensed, and (ii) to explain why the cloud should shed its spin angular momentum in discrete amounts, in isolated gas rings, rather than in a spatially continuous disc. There is also a third serious difficulty with the Laplace hypothesis, first noted by Fouche (1884): at the end of its radial contraction, the early sun should have been left spinning with a period of only 3 hours. The observed present-day solar period is 25 days. Fortunately, this problem is no longer a threat to the reconstruction of the Laplace hypothesis, as we point out in Section 2.3.

**2.2 A natural explanation for the Titius-Bode law**

The laws of conservation of total mass and angular momentum may be used to obtain the ratio of the orbital radii $R_n$ of the system of gas rings. These are numbered with index $n = 0, 1, 2, 3 \ldots$ :

$$\frac{R_n}{R_{n+1}} = \left(\frac{M_{n+1}}{M_n}\right)^3 \cdot \left(\frac{f_{n+1}}{f_n}\right)^2 \cdot \left[1 + \frac{m_{n+1}}{M_{n+1} f_{n+1}}\right]^2. \tag{1}$$

Here $M_{n+1}$ and $f_{n+1}$ denote the mass and moment-of-inertia factor of the PSC material that is interior to the equatorial cylinder of radius $R_e = R_{n+1}$ and $m_{n+1}$ is the mass of cloud material external to that cylinder. These cylinders have axes coincident with the spin axis of the PSC. Next, in deriving equations (1), it is assumed that $m_{n+1} \ll M_{n+1}$. This condition enables the so-called atmospheric approximation, Prentice (1978). The PSC spin angular velocity $\omega_{n+1}$ at the moment of detachment of the $(n+1)^{\text{th}}$ ring is given by the Keplerian value $\omega_{n+1} = (GM_{n+1}/R_{n+1}^3)^{\frac{1}{2}}$. The atmospheric approximation is consistent with a very centrally condensed PSC. The condition $m_{n+1} \ll M_{n+1}$ implies that $M_{n+1} \simeq M_n$. Equations (1) then reduces to the approximate form

$$\frac{R_n}{R_{n+1}} \simeq \left[1 + \frac{m_{n+1}}{M_{n+1} f_{n+1}}\right]^2. \tag{2}$$

If the contraction of the PSC takes place homologously, meaning that both $f_n$ and $m_n/M_n$ remain constant, then $R_n/R_{n+1}$ is also constant. That is, the sequence of mean orbital radii of the gas rings $R_n$ $(n = 0,1,2,3,...)$ form a geometric sequence. This feature of the MLT offers a natural explanation of the roughly-geometric structure of the Titius-Bode law of planetary distances.

**2.3 Importance of magnetic fields in ridding the final spin angular momentum**

In the past 60 years, since the pioneering observations of Herbig (1957), it has been discovered that young stars of near-solar mass, known as classical T Tauri stars (abbrev. CTTS), are slow rotators. They have spin periods that are 10 – 100 times longer than the local break-up rotational period $\sim$ 3 hr, Hartman & Stauffer (1989), Gallet & Bouvier (2013). These stars have radii $\lesssim 5R_\odot$. They also possess very strong magnetic fields, Johns–Krull (2007). It is very reasonable, therefore, to assume that during the final stages of gravitational contraction of the PSC, say for equatorial radii $R_e \lesssim 15R_\odot$, that magnetic fields enable the PSC to give up most of its residual spin angular momentum. That is, magnetic fields allow the PSC to acquire a final spin period characteristic of a CTTS, Prentice (in preparation, 2019). I note here that the loss of angular momentum through magnetic braking comes about through the coupling of the native magnetic field of the TT star with a surrounding disc of ionized gas, Königl (1991). The loss is very effective if there is a radial gap between equator of the star and the inner edge of the ionized disc, Hoyle (1960), and if this inner edge lies beyond the so called corotation radius, Ustyugova et al. (2006).

**3 SUPERSONIC THERMAL CONVECTION AS THE THEORETICAL FOUNDATION OF THE MLT**

It is proposed that the process of shedding gas rings comes about through the action of very powerful thermal convection that is generated by the release of gravitational energy from the radial contraction of the protosolar cloud (PSC). Strong convective currents of speed $v_t$ and mass density $\rho_t$ create a radial turbulent stress $p_{\text{turb}} = \langle \rho_t v_t^2 \rangle$ which, if the speeds are supersonic, can greatly exceed the local gas pressure $p_{\text{gas}} = \rho \Re T/\mu$. Here $\rho$ and $T$ are the local gas density and temperature, respectively, $\mathfrak{R}$ is the gas constant and $\mu$ is the mean molecular weight. This latter quantity depends on both the chemical composition and the temperature of the gas. In the absence of rotation and in the deep interior of the cloud, where the thermal convection is nearly adiabatic (see below), the turbulent stress has the form $p_{\text{turb}} = \beta\rho(r)GM(r)/r$, Prentice (1973), Prentice & Dyt (2003). Here $M(r)$ is the mass interior to radius $r$ and $\beta \cong 0.1$ is a constant called the turbulence parameter.

Prentice and Dyt (2003) produced a computational simulation of thermal convection in a model 2-dimensional, gravitational-stratified atmosphere that is heated strongly from below. They found that as all vertical motion ceases at the upper boundary, the uppermost layers of the atmosphere are strongly superadiabatic. That is, the vertical temperature gradient steepens sharply as the top of the gas layer is approached in order that the heat can be outwardly transported by thermal diffusion. A steep density inversion accompanies this event. Aside from the onset of superadiabaticity in the topmost layers, the convection in the remainder of the modelled atmosphere exists in near-adiabatic equilibrium.

**3.1 Defining the polytropic structure of the Protosolar Cloud**

The physical thermodynamic structure of the PSC is now assumed to parallel that of the 2-D simulated atmosphere. Let $r_0$ define the radius of the adiabatic core of the equivalent non-rotating PSC, whose surface radius $r_s$ always equals the polar radius $R_p$ of a rotating cloud, irrespective of the degree of rotation. This core is modelled as a polytropic gas in adiabatic convective equilibrium. In the outer regions of the core, where the gas is undissociated, the polytropic index is $n_{\text{ad}} = 2.342$ (see APPENDIX). In the inner region of the core, where the gas is dissociated or ionized, we have $n_{\text{ad}} = 1.500$. In the superadiabatic mantle surrounding the core, namely for radii $r$ such that $r_0 < r < r_s$, the polytropic index $n_{\text{sup}}$ satisfies the condition $n_{\text{sup}} < n_{\text{ad}}$. Now the polytropic relation that connects $\rho$ and $T/\mu$, namely $\rho \propto (T/\mu)^{n_{\text{sup}}}$, means that the local gas pressure behaves as $p_{gas} \propto (T/\mu)^{1+n_{sup}}$. Since physically both $p_{gas}$ and $T/\mu$ decrease with increasing radius, it is necessary that $n_{sup}$ also satisfy the additional condition $1 + n_{sup} \geq 0$. For the sake of simplicity, we set $n_{sup}$ equal to its most superadiabatic value, namely that $n_{sup} = -1$. This means that $\rho \propto \mu/T$ and hence that a density inversion takes place as the photosurface of the PSC is approached, in line with the simulated atmospheric study of Prentice & Dyt (2003). The mean molecular weight $\mu$ of the gas in the superadiabatic layer and in the outer undissociated layer of the adiabatic core, is constant: $\mu = \mu_s = 2.355$ (APPENDIX).

### 3.2 Introducing the 3 main parameters which control the structure of the PSC: $\beta_0$, $F_0$ and $\theta_s$

The turbulence parameter $\beta$ in the adiabatic core is assumed to be uniform and assigned the value $\beta_0$. We introduce a dimensionless temperature variable $\theta = T\mu_c/T_c\mu$, where the subscript 'c' refers to central values. The value of $\beta$ in the outer superadiabatic layer is assumed to decline linearly to zero at the photosurface (denoted with subscript 's'), where all heat escapes freely into space. That is, in the outer layer we set $\beta = \beta(\theta)$ where $\beta(\theta) = \beta_0 \cdot (\theta - \theta_s)/(\theta_0 - \theta_s)$. Here $\theta_0 = T_0\mu_c/T_c\mu_0$ and $\theta_s = T_s\mu_c/T_c\mu_s$. It is convenient to define a superadiabatic layer temperature-contrast factor $F_0$, namely $F_0 \equiv \theta_0/\theta_s$. This quantity remains invariant during the homologous phase of the PSC's contraction. Rotation is included in the model using the atmospheric approximation, Prentice (1978). This step is valid since the internal mass distribution of the PSC is very centrally condensed. It is also supposed that the interior of the PSC rotates uniformly as a result of turbulent viscosity arising from the very strong convection. For a fully rotating cloud, the angular velocity $\omega_e$ at the equator equals the Kepler value, namely $\omega_{Kep} = (GM_e/R_e^3)^{0.5}$. Here $R_e$ is the equatorial radius and $M_e$ the mass of the PSC. $M_e$ is defined to be the mass interior to the equatorial cylinder of radius $R_e$ whose axis is parallel to the axis rotation of the cloud, as noted in Section 2.2. The degree of rotation of the PSC is measured by the rotational parameter $\Theta_e = \omega_e^2 R_e^3/GM_e$. The equatorial radius $R_e$ and polar radius $R_p$ of the PSC are related by the equation $R_e = \left(1 + \frac{1}{2}\Theta_e\right)R_p$. For a fully rotating cloud we have $\Theta_e = 1$ and $R_e = 1.5R_p$.

**4 QUANTIFYING THE PHYSICAL STRUCTURE AND THERMAL EVOLUTION OF THE PROTOSOLAR CLOUD**

In order to quantify the physical structure and evolution of the contracting protosolar cloud and its family of shed gas rings, it is necessary to specify the values of the controlling parameters $\beta_0$, $F_0$ and $\theta_s$. The turbulence parameter $\beta_0$ is the motor behind the MLT. The shedding of a discrete system of gas rings, at mean orbital radii $R_n$ ($n = 0,1,2,3 \dots$), does not occur unless $\beta_0 > 0.05$, Prentice (1978). For $\beta_0 \gtrsim 0.1$, the changes in the physical structure of the PSC due to increased turbulent stress level off. Quantities such as the polar axial moment-of-inertia factor $f_e$ plateau to asymptotic values. The choice of $\beta_0$ is thus secondary to the choice of $F_0$ and $\theta_s$.

**4.1 Selection of the dimensionless photosurface temperature parameter $\theta_s$**

The photosurface temperature parameter $\theta_s$ controls the geometric spacing ratio $R_n/R_{n+1}$ of the mean orbital radii of adjacent gas rings. This parameter is chosen so that geometric mean of the spacing ratios of the 6 rings that are shed between the orbits corresponding to Jupiter and Mercury, including a gas ring at each of these endpoints and one for Ceres, leads to an exact match for the observed mean geometric spacing ratio of the adjacent present-day planetary orbital radii $R_{n,\text{p}}$ from Jupiter to Mercury. The numbering sequence $\{n\}$ for the system of gas rings starts with $n = 0$ for Quaoar, $n = 1$, for Neptune, etc. Thus $n = 4$ denotes the Jupiter gas ring and $n = 9$ the ring for Mercury. The initial mean orbital radius $R_n$ of a given gas ring differs from the present orbital distance $R_{n,\text{p}}$ of the corresponding planet that condenses from it because of a secular expansion arising from the loss of the mass of the PSC during its contraction to its final polar radius $R_{p,\text{fin}} = R_\odot$. The mass of the PSC after detachment of the $n^{\text{th}}$ gas ring is $M_n$. The final mass is $M_\odot$. Conservation of orbital angular momentum leads to the equation

$$R_{n,p} = R_n \cdot (M_n/M_\odot). \tag{3}$$

The geometric mean of the observed orbital distance spacing ratios from Jupiter to Mercury is

$$\langle R_{n,p}/R_{n+1,p}\rangle_{obs} = [R_{4,p}/R_{9,p}]^{0.2} = 1.6815. \tag{4}$$

The parameter $\theta_s$ is obtained by simultaneously solving Equations (1) and (4) for each given value of $\beta_0$ and $F_0$. This results in solving the single equation:

$$\frac{R_{4,p}}{R_{9,p}} = \left[\frac{f_9}{f_4} \cdot \frac{M_9}{M_4}\right]^2 \cdot \prod_{i=5}^{9}\left(1 + \frac{m_i}{M_i f_i}\right)^2. \tag{5}$$

### 4.2 Using Mercury's high metal content to determine the parameters $\beta_0$ and $F_0$

It has long been recognised that the end-member planet Mercury is a valuable compositional marker that can be used to calibrate the thermodynamic state of the PSC, Lewis (1972, 1974). The observed high mean density of Mercury, namely 5.432 ± 0.003 g/cm³, Smith et al. (2012), Perry et al. (2011), implies a very high metal mass content. To estimate the metal mass fraction $X_{\text{metal}}$, a thermally-evolved, present-day structure model of Mercury has been constructed. It is assumed that all of the metal (Fe, Ni, Cr, Co and V) is separated from the rock into a central core. Modelling the thermal evolution of Mercury to present age, viz. 4560 Myr, the inner ~70% of the mass of the metal core is solid alloy and the outer portion is liquid metal (Prentice, 2008, 2011). The core is surmounted by a metal-free rock mantle that consists mostly of calcium and aluminium silicates. Mercury's mean density can be accounted for if $X_{\text{metal}} = 0.7096$.

Now it readily follows from Table 2 of the Appendix that the mass fraction of Fe, Ni & Co in the protosolar gas relative to the total possible mass of condensable chemical compounds, including all rocky oxides, sulphides and halides but excluding $H_2O$, $NH_3$ and $CH_4$, is ~0.271. Some physical or chemical process must, therefore, have 'led to the high ratio of metals to silicates in Mercury', Solomon et al. (2007). A central pillar of the gas ring condensation model is that Mercury's high metal content is due to a gas phase process of metal/silicate fractionation. This process was first quantified by Grossman (1972) and Lewis (1972) in their landmark studies of chemical condensation in the solar nebula disc. It has been successfully applied to quantify the expected metal mass fraction in the condensate that forms in protosolar gas rings, Prentice (1991, 2008). The nub of this metal/silicate fractionation mechanism is that if – as we show later in Section 5 – the gas pressure $p_9$ on the mean central orbit of the Mercurian gas ring ($n = 9$) is sufficiently high, i.e. of order $p_9 = 0.2$ bar, then the condensation temperature of metal alloy, namely $T_{metal} \cong 1730$ K, is ~90 K higher than that of the MgSiO$_3$–Mg$_2$SiO$_4$ silicates, namely $T_{mag-sil} \cong 1640$ K. If, at the same time, the gas ring temperature $T_9$ is only slightly less than $T_{mag-sil}$, so that the quantity $\Delta T_1 = T_{mag-sil} - T_n$ ~3 K, then most of the potential magnesium silicate condensate remains in the vapour phase, whilst the metals are all strongly condensed. In this circumstance the metal mass fraction $X_{metal}$ of in the condensate ~0.71, in line with the value deduced from Mercury's high mean density.

The quantity $\Delta T_2 = T_{metal} - T_{mag-sil}$ depends on the gas ring central-orbit pressure $p_9$. This quantity and hence the metal mass fraction $X_{metal}$ depends very strongly on the physical depth of the superadiabatic layer of the PSC, as specified by the temperature contrast factor $F_0$. Now $X_{metal}$ depends only weakly on the choice of $\beta_0$. Detailed calculations show that the maximum value of $X_{metal} = X_{metal}(F_0, \beta_0)$ with respect to the variable $F_0$, namely $X_{metal,max}(\beta_0)$, barely changes with respect to $\beta_0$ for $\beta_0 \gtrsim 0.1$. It is a remarkable circumstance that for the protosolar elemental abundances that have been adopted in this paper (APPENDIX), $X_{metal,max}(\beta_0)$, has almost exactly the same value as that, namely 0.7096, that is needed to explain Mercury's high mean density of 5.432 g/cm$^3$. As $\beta_0$ changes from 0.112 to 0.115, $X_{metal,max}(\beta_0)$, drops only slightly from 0.7102 to 0.7090. Choosing $\beta_0 = 0.1135$, $X_{metal} = 0.7096$ is achieved by selecting $F_0 = 9.0734$.

Summing up, we select the mid-range value 0.1135 for $\beta_0$ and choose $F_0 = 9.0734$, so ensuring the metal mass fraction for Mercury which accounts for this planet's mean density. Having thus fixed $\beta_0$ and $F_0$, the parameter $\theta_s$ is adjusted so that the geometric mean of the spacing ratios of the central orbit radii of adjacent gas rings, namely $\langle R_n/R_{n+1}\rangle_{\text{geom}}$, leads to an exact match for the observed geometric mean of the orbital radii of adjacent planetary pairs, from Jupiter to Mercury, including Ceres, as described in Section 4.1. The net result of this computation is that $\theta_s = 0.0023231$.

The initial mass $M_e(0)$ of the PSC is chosen so that the final cloud mass $M_{\text{fin}}$, after all equatorial mass shedding has ceased, is $M_{\text{fin}} = M_\odot$. Now $M_e(0)$ depends on the initial equatorial radius $R_e(0)$ of the fully rotating cloud. This is chosen to be $R_e(0) = 10{,}000\, R_\odot$, so accommodating a first shed gas ring at the orbit of Quaoar. For this choice of $R_e(0)$, we find $M_e(0) = 1.21775 M_\odot$. From Equation (3) it follows that the secularly-expanded present-day value of $R_e(0)$ is $R_{e,p}(0) = 12{,}177.5 R_\odot$. For the mean present-day semi-major orbital distance of Quaoar, I have used the orbital elements given in the JPL Solar System Dynamics planetary ephemeris DE405, generated on 2013-Apr-13. The most recent planetary ephemeris DE431, generated on 2018-Mar-23, provides a slightly larger value semi-major axis for Quaoar, namely $R_{0,p} = 9{,}375 R_\odot$. As this value differs by less than the 1% from the DE405 value, the latter value is retained in this paper.

**4.3 Modelling the gravitational contraction and thermal evolution of the PSC**

The PSC contracts nearly homologously through the dimensions of the present planetary system as long as each of the parameters $\beta_0$, $F_0$ and $\theta_s$ remains constant. A modest departure from uniformity comes about through the steady increase in the fractional mass dissociated hydrogen in inner portion of the adiabatic core . The internal structure and spatial temperature distribution of the PSC is obtained by solving the equation for gravitational equilibrium, Prentice (1973, 1978). The temperature $T_e$ at the equator of the fully-rotating cloud is given by the equation

$$\Re T_e(R_e) = \tfrac{3}{2}\left(\frac{1}{1+n_1}\right)\cdot\left(\frac{\mu_s\theta_s x_s}{I_s}\right)\cdot\left(\frac{GM_e}{R_e}\right),$$

giving

$$T_e(R_e) = 1.376\times 10^7 \cdot \left(\frac{\mu_s\theta_s x_s}{I_s}\right)\cdot\frac{M_e}{M_\odot}\cdot\frac{R_\odot}{R_e}. \qquad (6)$$

Here $n_1 = 1.5$ is the polytropic index of the dissociated core, $\mu_s = 2.355$ is the mean molecular weight of the undissociated gas at the surface of the PSC and $x_s = R_s/R_{\text{Emd}}$ is the dimensionless radius of the equivalent, non-rotating ($\Theta_e = 0$) cloud of radius $R_s = R_p = \tfrac{2}{3}R_e$, expressed in units of the Emden radius $R_{\text{Emd}}$. This latter quantity is given by $R_{\text{Emd}}^2 = (1+n_1)\Re T_c/(4\pi\mu_c G\rho_c)$, where $T_c$ and $\rho_c$ are the central temperature and density. Also, $I_s$ is the dimensionless mass of the PSC, expressed in units of the Emden mass $M_{\text{Emb}} = \left(\frac{4\pi}{3}\right)\rho_c R_{\text{Emd}}^3$.

During the contraction of the PSC from the initial orbital radius of the Quaoar gas ring, namely $R_0 = 7649 R_\odot$, to that of Mercury, namely $R_9 = 76.69 R_\odot$, the equatorial temperature $T_e$ increases with decreasing radius $R_e$ closely as $T_e \propto 1/R_e^{0.9}$. At a transition radius $R_{\text{tran}} = 30 R_\odot$, $T_e(R_{\text{tran}}) = 3661$ K and the residual cloud mass is $M_{\text{tran}} = 1.049795 M_\odot$. As the final temperature $T_{\text{fin}}$ corresponding to the final equatorial radius $R_{e,\text{fin}} = 1.5 R_\odot$ is estimated to be of order 5000 K, Ezer & Cameron (1965), it is clear that the homologous contraction of the PSC cannot continue unchecked after $R_e$ has shrunk below the transition radius $R_{\text{tran}}$. For $R_e < R_{\text{tran}}$, it is therefore proposed that the surface temperature parameter $\theta_s$ declines smoothly in a manner so that the equatorial surface temperature $T_e$ levels off to the final chosen value $T_{\text{fin}} = 5000$ K. Defining $y_e = \log_e\!\left(\frac{R_e}{1.5 R_\odot}\right)$, the simplest analytic description for $T_e(R_e)$ in the radial interval $1.5 R_\odot < R_e < 30 R_\odot$ is:

$$T_e(R_e) = T_s(x_e) \equiv 5000\cdot\exp(c_2 y_e^2 - c_3 y_e^3),\ c_2 = 0.119392,\ c_3 = 0.051446. \qquad (7)$$

We assume that the reduction in $\theta_s = T_s\mu_c/T_c\mu_s$ for $R_e < R_{\text{tran}}$ is brought about by a steady decline in the turbulence parameter $\beta$, from $\beta_0 = 0.1135$ to $\beta = 0$, according as

$$\theta_s(\beta) = \theta_s(\beta_0) \cdot \frac{1}{2}\left(1 + \frac{\beta}{\beta_0}\right), \quad \theta_s(\beta_0) = 0.0023231. \tag{8}$$

We also assume that as $\beta \to 0$ the degree of superadiabaticity of the outer layer of the PSC, as measured by the polytropic index $n_{\text{super}}(\beta)$, relaxes linearly to the adiabatic value $n_1 = 1.5$ for ionized gas according as

$$n_{\text{super}}(\beta) = \frac{5}{2} \cdot \left(1 - \frac{\beta}{\beta_0}\right) - 1 \tag{9}$$

I note here that during the final stage of contraction of the PSC all of the gas in the outer superadiabatic layer is fully dissociated and ionized, so $n_{\text{super}}(0) = 1.5$.

The radius $R_e$ of the PSC for radii $R_e < R_{\text{tran}}$ is then found by simultaneously solving equations (6) and (7) for each for each selected value $\beta < \beta_0$.

For equatorial radii $R_e > R_{\text{tran}} = 30R_\odot$, the value of $R_e$ is obtained by selecting a value for the dimensionless radius $x_0 = r_0/R_p$ of the inner dissociated core of the equivalent non-rotating PSC, expressed in Emden length units. Imposing the condition that the mass fractions of atomic and molecular hydrogen are equal at that point provides a unique algorithm for solving for $x_0$, Prentice (1978). The values of $x_0$ are prescribed by a contraction index $J$ defined by the equation $x_0 = x_0(J) = 0.05(J - 1)$, with $J = J_{\text{min}}, \ldots J_{\text{max}}$. The computed value of $R_e$ for each $J$ is denoted by $R_e[J]$. For the prescribed values of the parameters $\beta_0$, $F_0$ and $\theta_s$ given above, we find $R_e[52] = 10{,}005.8R_\odot$. This point lies just outside the chosen starting size for the PSC, namely $R_e(0) = 10{,}000R_\odot$. The cloud radius $R_e$ shrinks to $29.2R_\odot$ for $J = 153$. This is just inside the $\beta = \beta_0$ transition point, where $R_{\text{trans}} = 30R_\odot$.

## 4.4 Quasi-hydrostatic thermal equilibrium and the rate of contraction of the PSC

In order for the PSC to achieve a Kelvin-Helmholtz thermodynamic equilibrium, it is necessary that the modulus of the gravitational potential energy of the cloud (minus the sum of the thermal, turbulent and rotational energies), namely $|E_{\text{grav}}|$, exceed the sum of the dissociation energy of H$_2$ and ionization energies of H$_1$ and He, namely $E_{\text{diss-ion}}$. Unfortunately, this condition is not satisfied until the equatorial radius $R_e$ has shrunk to the Kelvin-Helmholtz value $R_{\text{K-H}} \sim 150 R_\odot$. This value lies inside the initial mean orbital radius of the Earth ($\sim 190 R_\odot$). It has been proposed, therefore, that in order to achieve a quasi-static thermal equilibrium for cloud sizes larger than $R_{\text{K-H}}$, a small fraction of the PSC initial cloud mass $M_e(0)$ collapses freely to near-stellar size to form a compact central core, Prentice (1978). A core of radius $r_{\text{core}} = 3 R_\odot$ and mass $m_{\text{core}} = 0.0392 M_\odot$ releases sufficient gravitational energy to stabilize the in-fall of the remainder of the cloud mass, for an assumed initial cloud radius $R_e(0) = 10,000 R_\odot$. The initial PSC mass, namely $M_e(0) = 1.21775 M_\odot$, is chosen so that the final cloud mass, after all mass shedding has ceased, is $M_{\text{fin}} = M_\odot$. This event occurs near radius $\sim 5 R_\odot$, as mentioned earlier in Section 2.3.

### 4.4.1 Quantification of the rate of radial contraction for equatorial radii $R_e \leq R_{\text{K-H}}$

The quantification of the rate of radial contraction of the PSC for radii $R_e \leq R_{\text{K-H}}$ is well understood, Prentice (1978). The PSC contracts at a the rate at which the luminosity at the photosurface, namely $L_{\text{ph}} = 2.8113 \pi \sigma_s R_e^2 T_e^4$, radiates into free space the change $\Delta E_{\text{total}}$ in total energy $E_{\text{total}}(R_e) = E_{\text{diss-ion}} - |E_{\text{grav}}|$ between the cloud radii $(R_e + \Delta R_e)$ and $R_e$. Here $\Delta E_{\text{total}}(R_e)$ equals the sum of the ionization and dissociation energies, namely $E_{\text{diss-ion}}$, and the negative gravitational energy of the PSC (less the thermal, turbulent and rotational energies), namely $-|E_{\text{grav}}|$. Complete hydrostatic equilibrium is achieved only if $E_{\text{total}}(R_e) \leq 0$. The time $\Delta t$ taken to contract from radius $R_e + \Delta R_e$ to $R_e$ is given by

$$\Delta t = [E_{\text{total}}(R_e + \Delta R_e) - E_{\text{total}}(R_e)]/L_{\text{ph}}. \tag{10}$$

Numerical values: For the cloud radius index $J = 133$: $R_e[J] = 148.443 R_\odot$, $T_e[J] = 865.7$ K and $E_{\text{diss-ion}}/|E_{\text{grav}}| = 1.003$. For $J = 52$: $R_e[J] = 10{,}006.8 R_\odot$, $T_e[J] = 20.90$ K and $E_{\text{diss-ion}}/|E_{\text{grav}}| = 2.691$. The index $J = 133$ thus marks the start of the true Kelvin-Helmholtz contraction, giving $R_{\text{K-H}} = R_e[133] = 148.443 R_\odot$. The speed of radial contraction of the PSC at that point is then found from equation (10). We have $v_e[153] \cong ((R_e[152] - R_e[153])/\Delta t$. The computed PSC model now yields

$$\left[\frac{dR_e}{dt}\right]_{J=153} = 0.01014\ R_\odot/\text{year}. \qquad (11)$$

4.4.2 Quantification of the rate of radial contraction for equatorial radii $R_e > R_{\text{K-H}}$

After the initial PSC has become energetically stabilized by the formation of a small 'cometary' core, we may assume that the contraction proceeds quasi-statically at the rate controlled by the photospheric luminosity $L_{\text{ph}}$. The computer modelling shows that the principal change in total energy $E_{\text{total}}(R_e)$ with decreasing $R_e$ comes about mostly though the radial change in $-|E_{\text{grav}}|$. The rate of change of $E_{\text{diss-ion}}$ is small compared to that of $-|E_{\text{grav}}|$. Hence

$$\frac{d}{dt}(E_{\text{total}}) \approx \frac{d}{dt}(E_{\text{grav}}) \propto \frac{1}{R_e^2}\frac{dR_e}{dt}$$

noting that $E_{\text{grav}} \propto GM_e^2/R_e$. Next, the numerical modelling (see above) shows that in the interval from $R_{\text{K-H}}$ to $R_e(0)$, we have $T_e \propto 1/R_e^{-0.9}$ and hence that $L_{\text{ph}} \propto R_e^{-1.6}$. Equating $dE_{\text{total}}/dt$ with $L_{\text{ph}}$, leads to the result that for the radius interval $R_{\text{K-H}} \leq R_e \leq R_e(0)$

$$[dR_e/dt]_{J \leq 153} = 0.01014(R_e/R_{\text{K-H}})^{0.4}\ R_\odot/\text{year}. \qquad (12)$$

Equation (12) may now be integrated to give the equatorial radius $R_e$ as a function of the time $t$ (unit: year) that elapses from the starting radius $R_e(0) = 10{,}000 R_\odot$. We have

$$R_e(t) = R_e(0)\cdot[1 - t(\text{yr})/(3.051\times 10^5)]^{5/3} \quad \text{with} \quad 150 R_\odot \lesssim R_e \leq 10{,}000 R_\odot \qquad (13)$$

The remarkable conclusion that emerges from Equation (13) is that it takes approximately only $2.8 \times 10^5$ years for the protosolar cloud to contract from the initial radius $10{,}000 R_\odot$, to well inside the orbit of the Earth. The time for contraction to the transition radius $30 R_\odot$ is $3.0 \times 10^5$ years. These calculations suggest that the process of planet formation took place on the very short timescale of order $3 \times 10^5$ years.

**5 NUMERICAL RESULTS: CHEMICAL CONDENSATION OF THE PLANETARY SYSTEM**

We use the quantified model for the thermal evolution of the PSC that has been established in Section 4 to compute the expected bulk chemical composition of the solids that form within the family of shed gas rings. The first part of this process is to compute the carbon chemistry for each of the gas rings. That is, we compute how carbon is distributed between the 4 possible compound states, namely gaseous CO, $CO_2$, and $CH_4$ and solid carbon, denoted as C(s) (i.e. graphite). The 2nd part of the computation process is to determine the suite of other chemical compounds that can form within each gas ring. This latter process depends on the temperature and mean orbit gas pressure of each of the protosolar rings as well, of course, on the elemental abundances of the protosolar gas. The elemental abundances are given in the APPENDIX to this paper.

**5.1 Carbon chemistry of the protosolar cloud**

In 2 earlier papers, Prentice (1990, 1993a) has laid out the theoretical foundations for computing the carbon chemistry within the PSC. This theory is an extension of a landmark study of carbon chemistry in the solar nebula, Lewis & Prinn (1980). The main point made by Lewis & Prinn is that as the temperatures $T_{\text{neb}}$ and pressures $p_{\text{neb}}$ in the outer solar nebula are very low, typically $T_{\text{neb}} < 100$ K and $p_{\text{neb}} < 10^{-6}$ bar, the production of $CH_4$ and C(s) via gas phase chemistry is kinetically inhibited. That is, in the absence of a mechanism for the conversion of CO (and $CO_2$) to $CH_4$ and C(s), carbon will remain as CO and $CO_2$, as inherited from the interstellar medium. To get around this problem, Lewis & Prinn suggested that the production of $CH_4$ and C(s) can take place if we take into account, firstly, the possibility of catalysed reactions on the surfaces of metallic grains consisting of pure Fe, Ni and Co. The conditions for these reactions to occur exist deep inside the solar nebula, where the temperature exceeds 400 K and the pressures are larger. Secondly, as long as a mechanism exists for transporting the reaction products of the catalytic chemistry to the outer regions of the nebula, we can explain the presence of $CH_4$ and C(s) in outer solar system bodies.

Prentice (1990, 1993a) noted that the MLT meets both of the criteria stipulated by Lewis & Prinn (1980). Firstly, the temperatures and pressures needed for surface-catalysed chemistry do exist within the warm dense interior of the PSC. Secondly, as powerful thermal convection is the basic ingredient of the MLT, responsible for the shedding of discrete gas rings, catalysed reaction products are swiftly advected to the photosurface of the PSC. Once those products reach the surface and become incorporated into the non-turbulent equatorial gas rings, any further processing of CO is quenched because of the low temperatures and pressures there.

The details of the theory for the catalysed production of $CH_4$ and C(s) on the surfaces of small metallic grains consisting of pure Fe, Ni and Co are given in Prentice (1990, 1993a). It is sufficient here to note that the rates of catalytic production of $CH_4$ and C(s) depend principally on (i) the radius $a_{gr}$ of the metallic grains, and (ii) the catalytic synthesis temperature $T_{cat}$. The values of $a_{gr}$ and $T_{cat}$ are chosen: (i) to yield a mean mass density of the condensate at the orbit of Quaoar of 1.73 g/cm³, and (ii) to ensure that all C(s) production ceases by the time the PSC has contracted to roughly midway between the orbits of Ceres and Jupiter, namely about ~800 AU. I assume that the majority of the main belt asteroids were formed stony, free of carbon. Both conditions (i) and (ii) are satisfied with the choices $a_{gr} = 5 \times 10^{-5}$ cm and $T_{cat} = 465$ K. The mean density 1.73 g/cm³ is derived from models for the formation of the Pluto-Charon binary, Prentice (2015). The other parameters that form part of the catalytic reaction model are the activation energy $E_{act}$ for grain surface chemistry and the free metal abundance concentration $F_{met,free} = \Sigma_{i=1}^{3} X_{i,free}/\rho_i$.

For $E_{act}$, I adopt 15 kcal/mole. From the APPENDIX, we find $F_{met,free} = 9.26 \times 10^{-5}$. In calculating $F_{met,free}$, I have removed all Fe, Ni, and Co that is tied up in sulphides. The formation of FeS, NiS and CoS proceeds for temperatures less than 700 K. Following Fegley & Prinn (1989), I assume that the interconversion of CO and $CO_2$, via the rapid water-gas shift reaction, continues until quenched at a temperature $T_{CO,CO_2}$ that is 50 K below $T_{cat}$. That is, I set $T_{CO,CO_2} = 415$ K.

The results of the computation for the synthesis of methane and graphite via surface–catalysed reactions are displayed graphically in Figure 1. The 4 curves show the distribution of carbon by number fraction between the 4 possible states of CO, $CO_2$, $CH_4$ and C(s). These numbers are plotted against the present-day equivalent orbital positions $R_{e,p}$ of the equatorial radius $R_e$ of the PSC, as explained in Section 4.1. That is, the abscissa variable in Figure 1 is $\log_{10}(R_{e,p}/R_\odot)$, where $R_{e,p} = R_e \cdot (M_e/M_\odot)$. The catalytic production model assumes that the contraction of the PSC takes place as a continuous function of the equatorial radius $R_e$. The time $t$ taken for the cloud to contract to radius $R_e$ can be readily found from Equation (13). At time $t = 0$, we have $R_{e,p}(0) = 12{,}177.5\ R_\odot$. The actual times $t_n$ for each of the protosolar gas rings corresponding to the present day orbital positions of the planets, including Ceres and Quaoar (KBOs), are given in Table 1. For convenience, it is assumed that all carbon is initially present as CO. This assumption has no effect on the solution for the carbon distribution for time $t > 0$.

**5.2 Computation of the bulk chemical compositions of the protosolar ring condensates**

The computation for the bulk chemical composition of the condensate that forms within the family of gas rings shed by the PSC requires an exact specification of the elemental composition of the protosolar gas. This is given in the APPENDIX. The key data used for its construction are:

1. Abundances of rock-like and metallic elements are the CI-chondrite abundances, Lodders et al. (2009).
2. Abundances of C, N and O are the solar photospheric values, Caffau et al. (2010), (2009) & (2008), respectively. These data yield a [C/O] molar ratio of 0.55.

3. Solar Ar abundance, Lodders (2008).
4. Solar H abundance is the helioseismological value, Basu & Anita (2004) & (2008).
5. Algorithm for deducing protosolar abundances from solar abundances taken from Asplund et al. (2009) and Turcotte & Wimmer-Schweingruber (2002).
6. Atomic weights, Eidelman et al. (2004).

Two important findings that come from the protosolar abundance determination are: (1) the bulk H mass fraction is X = 0.7163. The heavy element mass fraction (including C, N & O) is Z = 0.01604.

**5.3 The temperatures and chemical condensation sequence for the system of protosolar gas rings**

The heavy white diagonal curve in Figure 2 shows the temperature $T_e$ at the equator of the PSC at the moment of detachment of a local gas ring. The gravitational contraction and thermal evolution of the PSC is modelled as a continuous function of the equatorial radius $R_e$ of a fully rotating cloud, as specified by the rotation parameter $\Theta_e = 1$. All of the results are computed for the values of the controlling parameters $\beta_0 = 0.1135$, $F_0 = 9.0734$ and $\theta_s = 0.0023231$ established in Section 4.1. In Figure 2 (as for Figure 1) the temperature $T_e$ is plotted as a function of the present-day equivalent equatorial radii $R_{e,p}$, as explained in Section 4.1. The small open circles on the heavy locus mark the temperatures $T_n$ of the gas rings corresponding to the initial orbital locations $R_n$ of each of the 4 outer planets Jupiter, Saturn, Uranus and Neptune, as well as for the initial orbital distance of Quaoar. The KBOs are assumed to have condensed from that first gas ring. The temperature within each ring is assumed to be spatially uniform and equal to $T_n$.

The other curves in Figure 2 specify the condensation temperatures $T_{n,j}$ of the principal low-temperature chemical species $\{j\}$. Here we define $j$ as a species index, with $j = 0, 1, 2, 3, …$ . These species include the principal ices $H_2O$, $CO_2$, $CH_4$ and the clathrate hydrates of $CH_4$ and Ar. The condensation curves for the 4 principal low-temperature rock phases are also shown in the Figure. The reader is referred to Prentice (2015) for the graphic that displays the corresponding condensation temperature loci of the high-temperature rocky phases, including the metals. This latter graphic applies to the condensation of the terrestrial planets and the main belt asteroids, including Ceres.

The condensation temperature $T_{n,j}$ of a given chemical species $j$ is defined to be the value calculated for the partial pressure $p_{n,j}(0)$ on the central mean orbit of the gas ring. We have $p_{n,j}(0) = p_n(0) \cdot (\mu X_j/\mu_j)$, where $p_n(0)$ is the mean orbit total gas pressure, $\mu$ is the gas mean molecular weight, $X_j$ is the mass fraction of the species $j$ in the gas, prior to condensation, and $\mu_j$ is the species molecular weight. The formula for computing $T_{n,j}$ is given by

$$T_{n,j} = A_j/[B_j - \log_e(p_{n,j}(0))]. \tag{14}$$

Here $A_j$ and $B_j$ are well-defined thermodynamic constants. They are determined by the entropy and enthalpy of the gas. In the case of the condensation of $CO_2$ ice, for example, the data of Miller & Smythe (1970) give $A_j = 3143$, $B_j = 16.19$. For the KBO (Quaoar) gas ring, therefore, where $X_j = 4.81 \times 10^{-5}$ and $p_n(0) = 1.265 \times 10^{-9}$ bar (see Table 1), we obtain $T_j = 68.7$ K. The final step in the process is to calculate the actual mass fraction $X_{j,\text{cond}}$ of the total initial $CO_2$ vapour that condenses as ice. No condensation takes place unless $T_{n,j} < T_n$. In the case of the gas ring configuration of the Modern Laplacian Theory, the condensing mass fraction is by the equations

$$X_{j,\text{cond}} = [1 - (1 + \varphi_j) \cdot \exp(-\varphi_j)] \cdot X_j, \quad \varphi_j = A_j \cdot [1/T_n - 1/T_{n,j}]. \tag{15}$$

Table 1 gives the values of the mean orbital radii (initial and present-day), the temperatures $T_n$, and mean orbit gas pressures $p_n$ of the protosolar gas rings that are shed by the PSC in the outer solar system, from Jupiter to Quaoar, as well as the values for the gas rings shed at the orbits of Ceres and Mercury. The times $t_n$ of detachment from the PSC are shown, measured from time $t = 0$ when the initial equatorial cloud size is $R_e(0) = 10{,}000 R_\odot$. The residual mass $M_n$ of the PSC after the detachment of the $n$-th ring is also given.

Table 2 gives the normalized mass fractions $X_j$ of the principal bulk chemical condensates that form from each of the outer protosolar gas rings, as well as for Ceres. There are several points worth noting in Table 2. First, the production of elemental carbon achieves a maximum at the orbit of Saturn and has vanished altogether by the orbit of Ceres. Second, $CO_2$ ice can condense only at the orbits of Uranus and beyond. Third, the formation of $CH_4$ hydrate clathrate, namely $CH_4 \cdot 5¾H_2O$, precedes that of $CH_4$ ice. As $CH_4$ saturates all of the available $H_2O$ ice, no $H_2O$ ice is left over for the formation of Argon hydrate clathrate. Fourth, condensation of pure $CH_4$ ice occurs only at the orbit of Quaoar. That is, it is only the KBOs that contain pure $CH_4$ in their condensation mix. Lastly, I note that the rock is slightly hydrated as the result of the condensation of the hydrated species NaOH and KOH. These make up 1.2% of the rock mass mix. $Mg(OH)_2$ condenses interior to the orbit of Saturn. It makes up 27% of the Jovian rock mass. The last column in Table 2 gives the mean density $\rho_{\rm cond}$ of the condensate. It has been computed for the present-day blackbody temperature $T_{\rm bb}$ at each of the planetary orbits $R_{n,\rm p}$. For the KBOs I take $T_{\rm bb} = 35$ K. The mean density of the rock component of the KBO unprocessed condensate mix is 3.685 g/cm³.

## 6 THERMAL EVOLUTION AND PHYSICAL STRUCTURE OF ULTIMA THULE

The last step in this paper is to compute the predicted present-day internal structure, surface composition and appearance of Ultima Thule, in readiness for the New Horizons flypast. Prior to New Horizons, it appears that very little theoretical modelling has been published on the thermal evolution of Kuiper Belt objects that have mean radius in the range 10-20 km that is anticipated for Ultima. Moore et al. (2018) have drawn together much of the current knowledge. The most thorough thermal investigation of KBO objects having mean radius < 20 km is that of Prialnik et al. (2008). These authors chose rock and amorphous water ice as the 'bedrock' constituent. $CO_2$ ice was treated as a secondary constituent. $CH_4$, along with CO and $N_2$, were assumed to be present as vapour trapped in the $H_2O$ ice. The primary goal of Prialnik et al. (2008) was to investigate the influence of radiogenic heating on the reduction of porosity in amorphous water ice that comes about through the release of the trapped volatile gases. It is the release of these vapours, followed by their upward migration and subsequent possible refreezing at the surface, which may account for the observed preponderance of CO and $N_2$ ices in many comets. Of course, some of these volatiles may escape before freezing. In the present study CO and $N_2$ are not included as they are not part of the compositional mix. $CH_4$ is a natural condensate in the model, not a trapped gas.

Prialnik et al. (2008) have demonstrated that heat derived from the decay of the short-lived radionuclide radioactive isotope $^{26}$Al is sufficient to melt $H_2O$ ice in KBOs having mean radii as small as 10 km. It was Urey (1955) who first recognised that the primary source of heat in small solar system objects, such as Ultima Thule, is due to the decay of short-lived radionuclides.

**6.1 A computational model for the thermal evolution of the 2 single lobe components of Ultima**

Moore et al. (2018) suggest that Ultima may be a bi-lobate object of total size roughly $20 \times 35$ km. This data can be matched by 2 single contacting lobes having mean radii of 10 km and 7.5 km. For simplicity, I assume that each lobe is a perfect sphere. Of course, the true sizes of the lobes will not be known after the New Horizons flypast. The choice of radii is thus arbitrary at this stage.

From the point of view of thermal modelling, I have chosen two lobes that differ by 20% in radius. Lobe 1 has a radius of 10.5 km, Lobe 2 has radius 8.4 km. I model the thermal evolution of each lobe assuming that they are compositionally uniform and have initial bulk chemical composition the same as that derived in Section 5. One slight variation is that the initial slightly-hydrated rock is assumed to quickly dehydrate through radiogenic heating, so releasing the water associated with each of NaOH and KOH. If all of the NaOH and KOH is converted to form alkali feldspars, the new combined rock-graphite mass fraction is 0.5418, in place of 0.5431. The water ice mass fraction increases slightly from 0.1845 to 0.1858. Another minor change, concomitant with that above, is to assume that any $CH_4$ tied up as clathrate hydrate is released. This yields a total pure CH$_4$ ice fraction of 0.0513. The mean mass density of the bulk condensate at 35 K is now 1.730 g/cm³. This density is computed for 1 bar pressure.

Next, the computational code EVOL9CTC. FOR that has been developed for modelling the thermal evolution of the satellites of Jupiter and Saturn, Prentice (2001, 2006), has been extended to include the heat released by the decay of short-lived radionuclide $^{26}$Al. Up until now, only the heat derived from the decay of the long-lived radionuclides of K, U and Th has been included.

6.1.1 Data for modelling radiogenic heating due to decay of $^{26}$Al

The following $^{26}$Al data have been adopted for the EVOL9CTC. FOR upgrade:
1. Isotopic number concentration: [$^{26}$Al/$^{27}$Al] = $5 \times 10^{-5}$
2. Half-life decay time: $7.17 \times 10^5$ yr.
3. Specific isotopic heat production rate: 0.4554 W/kg
4. Mass fraction of total Al in anhydrous Ultima rock: 0.01179

**6.2 Numerical results: thermal evolution**

Figures 3 and 4 show the temperature evolution $T(r,t)$ for each of the 2 single lobes of Ultima. The temperature is plotted as a function of the fractional internal radius $x = r/R_{\text{lobe}}$ for several representative times $t$ (unit: Myr). The surface temperature is kept constant during the thermal evolution, namely $T_{\text{surf}} = 35$ K

6.2.1 Thermal evolution of Lobe 1: radius 10.5 km

We see in Figure 3 that the temperature rises very quickly throughout the structure. The central temperature reaches a peak of 570 K at $4.0 \times 10^5$ yr. In the outer layers of the lobe, the temperatures peak sooner at $2.0 \times 10^5$ yr. By this time the temperature exceeds the melting point of CH$_4$ ice, namely $T_{\text{m}}(\text{CH}_4) = 90.7$ K, throughout the inner 66% of the lobe mass. For computational simplicity, it is assumed that the lobe remains chemically homogeneous throughout its thermal evolution. In this instance, the fractional mass $I(x)$ interior to fractional radius $x$ is given by the equation $I(x) = x^3$. Next, by the same time of $2.0 \times 10^5$ yr, the temperature exceeds the sublimation temperature $T_{\text{sub}}(\text{CO}_2)$ of CO$_2$ ice throughout the inner 39.3% of the lobe mass. Now $T_{\text{sub}}(\text{CO}_2)$ is a function of the local hydrostatic pressure $p$. An implicit equation for $T = T_{\text{sub}}(\text{CO}_2)$ (unit: K) as a function of $p$ (unit: bar) can be derived from empirical data given in the IUPAC International Thermodynamical Tables of the Fluid State Carbon Dioxide, Angus et al. (1976). I obtain

$$\log_{10}(p/\text{bar}) = 8.2240 - 1396.19/T - 0.4550\log_{10}(T) \qquad (16)$$

The $CO_2$ sublimation temperature is shown by the heavy dashed locus in Figure 3. For Lobe 1 the pressure $p$ at the 0.393 mass fraction point and time $t = 0.2$ Myr, when $T(r,t) = T_{\text{sub}}(\text{CO}_2) = 177.4$ K, is $p = 0.214$ bar. At time $t = 0.2$ Myr the temperature has risen above the melting point of $H_2O$, namely $T_{\text{m}}(\text{H}_2\text{O}) = 273.15$ K, throughout the inner 21.5% of the mass of Lobe 1.

6.2.2 Thermal evolution of Lobe 2: radius 8.4 km

The thermal evolution for the smaller lobe is similar to that of the 10.5 km radius lobe, except that the temperatures are everywhere lower and that the peak in temperature at each radius point occurs sooner. The central temperature rises to only 327 K and this happens at time $2.0 \times 10^5$ yr (curve not displayed). In the outer layers peak temperatures are reached at time $t = 1.5 \times 10^5$ yr when the temperature exceeds the melting point of $CH_4$ ice in the inner 53% of the lobe mass. The sublimation of $CO_2$ ice occurs within the inner 21% of the mass of the lobe and the melting of water ice takes place only near the very centre, within the inner 3.5% of the lobe's mass.

**6.3 Implications of the thermal evolutionary models for the expected present-day structure of Ultima Thule**

The above modelling confirms the earlier conclusion of Prialnik et al. (2008) that unless the radius of a KBO appreciably exceeds 10 km, extensive melting of water ice in the interior is very unlikely. Thus if Ultima Thule were to consist solely of water ice, in addition to rock, one would expect to see very little evidence of internal alteration of its physical structure taking place as a result of the melting of this ice, and especially so in the smaller of the two Ultima lobes. If, however, we are prepared to include $CH_4$ and $CO_2$ ices as significant mass components of the 'bedrock' bulk compositional mix, then a very different outcome may occur. We have found that if the radius of a lobe exceeds 10 km, melting of $CH_4$ ice takes place throughout a very large fraction of the interior. Now liquid $CH_4$ is extremely light, having mass density ~0.4 g/cm$^3$. Liquid $CH_4$ will therefore migrate immediately towards the surface, via porous voids. It should then refreeze in the cool outer layers of the object to form a methane-enriched outer shell. It is very reasonable to assume that being a rising liquid, rather than a vapour, very little of the $CH_4$ escapes into space.

It is the sublimation of $CO_2$ ice, however, that has the potential to produce a most profound alteration in the internal physical structure of the object. Provided that the radius of a lobe exceeds ~12 km, at least ~50% of the $CO_2$ ice will turn to vapour. The highly-buoyant $CO_2$ vapour should normally migrate swiftly to the surface through porous voids and escape into space. The escape of the $CO_2$ vapour is, however, unimpeded only if there remain open passage-ways to the surface. But since (1) the melting of $CH_4$ ice precedes the sublimation of $CO_2$ ice, and (2) as most of the buoyant $CH_4$ liquid is expected to refreeze at the surface, so forming a non-porous crystalline outer crust, the physical conditions now exist for the imprisoned and pressurized $CO_2$ vapour to erupt explosively through the upper layers of the globe, so disrupting the outer primordial crust of Ultima Thule. Much of the original store of $CO_2$ ice may therefore be lost. As the surface temperature is at most 35 K, a sizable portion of escaping $CO_2$ vapour may be cooled sufficiently to refreeze during its passage through the outer methane crust. The final outer crust of Ultima Thule may thus consist of a spatially jumbled mixture of $CH_4$ and $CO_2$ ices.

**6.4 A schematic model for the internal physical structure of a two-lobed Ultima Thule**

Figure 5 is a graphic which illustrates schematically how the physical structure of a two-lobed Ultima Thule may have changed as result of the combined actions of the early melting of $CH_4$ ice and the subsequent sublimation of $CO_2$ ice. The graphic has been constructed on the assumption that Ultima Thule today consists of 2 globes of radius 10 km and 7.5 km, as noted earlier. The model assumes that the entire original store of $CO_2$ ice sublimed as result of heat released through the decay of $^{26}$Al. If ~60% of the $CO_2$ is lost into space as vapour and ~40% remains to become part of the outer frozen methane-rich crust, the original object shrinks in physical size by ~5%. A lobe of initial radius 10.48 km shrinks to 10 km and a 7.86 km radius lobe shrinks to 7.5 km.

In constructing Figure 5, it is supposed that large-scale melting of water ice allows rock-graphite particles to settle towards the centre of each globe to form a dense core. The mass fraction of this core is 0.6221 if ~60% of the original store of $CO_2$ escapes from each lobe. As all radiogenic heat is now long gone, a uniform temperature of 35 K is expected everywhere in each present-day lobe. The mean density of both of the fully-differentiated lobes illustrated in Figure 5 is 1.734 g/cm$^3$. The central pressure of lobe 1 is 0.987 bar. It is an unusual quirk that as the density of frozen $CO_2$ at 35 K is 1.701 g/cm$^3$, the mean density of either lobe is not that much different from that (1.730 g/cm$^3$) of the initial undifferentiated bulk chemical mix for Ultima, prior to the loss of any $CO_2$.

## 7 PREDICTIONS FOR THE STRUCTURE AND SURFACE APPEARANCE OF ULTIMA THULE FOR THE NEW HORIZONS ENCOUNTER AND CONCLUSIONS

### 7.1 Predicted structure and surface appearance

It is predicted that as long as the 2 lobes of Ultima are sufficiently large, so that the heat released by the decay of the radioactive isotope $^{27}$Al plays a very significant role in the early thermal evolution of Ultima, the New Horizons spacecraft may discover that very little remains of the original surface appearance and topography of this KBO. All of the original crater record that was acquired during Ultima's accretion may have been erased by the early thermal activity of the $CH_4$ and $CO_2$ ices. The surface today may thus be nearly crater free. At small distance scales, the surface of Ultima Thule may have a 'gardened' appearance. It is predicted that a hybrid mix of pure $CH_4$ and $CO_2$ ices make up ~11% of the outer radial distance of Ultima's physical structure.

The top $CH_4$ becomes damaged by cosmic rays and ultraviolet radiation, so making Ultima Thule appear dark. $CO_2$ ice should be detected everywhere.

At larger distance scales, the surface is expected to be globally smooth and have a shape determined by gravity and rotational force, rather than material strength. This situation comes about as a result of the combined actions of (1) early melting and upward migration of liquid $CH_4$ to form a smooth outer crystalline shell, and (2) explosive escape of $CO_2$ vapour trapped beneath the outer non-porous $CH_4$ shell. I note here that A'Hearn et al. (2011) have suggested a similar explanation for the smooth waist profile of comet Hartley 2.

**7.2 A note of caution and finale**

The results of our modelling of the thermal evolution of a KBO bodies having radii of order ~10 km and ~8 km indicate that it is very unlikely that a bi-lobate object having lobes of this size will evolve to the smooth structure shown schematically in Figure 5. That is, if Ultima Thule is a bi-lobate object having overall size about $20 \times 35$ km, as suggested by the stellar occultation data, then it has most probably retained much of its original topography.

If Ultima Thule is discovered by New Horizons to be single lobe structure, perhaps having the shape of an oblate spheroid with mean radius about 15 km, then all of the calculations that were initiated by Prialnik et al. (2008) and expanded here to include a large native store of $CH_4$ and $CO_2$ ices, may bear fruit. I do hope so!

A shorter account of my predictions for Ultima Thule is given elsewhere, Prentice (2019).

**ACKNOWLEDGEMENTS**

The author thanks S. Morton, A. Craig and N. Porcellato for providing much technical support in producing the graphics in this paper, and P. Cally, J. Cashion, D. Collins, D. Cruikshank, M. Marshall and C. Morgan for helpful discussions. Revd. L. Thompson assisted in assembling the references. He gratefully acknowledges the support of the Monash University IT department, especially of A. Thorne and S. Ziemer, for maintaining the University's last remaining DEC Alpha machine.

**REFERENCES**

A'Hearn, M.F., & 33 others, 2011, Sci., 332, 1396.

Angus, S., Armstrong, B., de Reuck, K.M., Altunin, V.V., Gadetskii, O.G., Chapela, G.A. & Rowlinson, J.S., 1976, International Thermodynamic Tables of the Fluid State: Carbon Dioxide, (Pergamon Press: Oxford), p.380.

Asplund, M., Grevesse, N., Sauval, A.J., & Scott, P., 2009, Annu. Rev. Astron. Astrophys., 47, 481.

Auer, L.H., & Woolf, 1965, N.J., ApJ, 142, 182.

Babinet, M, 1861, Comptes Rendus Acad. Sci Paris, 52, 481.

Basu, S., & Anita, H.M., 2004, ApJ, 606, L85.

Basu, S., & Anita, H.M., 2008, Phys. Reports, 457, 2117.

Buie, M.W., & 12 others, 2017, AGU Fall Meeting, abstract #P13F-08.

Caffau, E., Ludwig, H.-G., Steffen, M., Ayres, T.R., Bonifacio, P., Cayrel, R., Freytag, B., & Piez, B., 2008, AA, 488, 1031.

Caffau, E., Ludwig, H.-G., Bonifacio, P., Farraggiana, R., Steffen, M., Freytag, B., Kamp, I., & Ayres, T.R., 2010, AA, 514, A92, I.

Caffau, E., Maiorca, E., Bonifacio, P., Farraggiana, R., Steffen, M., Ludwig, H.-G, Kamp, I., & Busso, B., 2009, AA, 498, 877.

Eidelman, S., & many others, 2004, Phys. Letts. B, 592, 1.

Ezer, D. & Cameron, A.G.W., 1965, Can. J. Phys., 43, 1497.

Fegley, B., & Prinn, R.G., 1989, in The Formation and Evolution of Planetary Systems, ed. H.A. Weaver & L. Danly, (Cambridge: CUP), p. 171.

Fernández, J.A., 1980, MNRAS, 192, 481.

Fouché, A., 1884, Comptes Rendus Acad. Sci. Paris, 99, 903.

Gallet, F., & Bouvier, J., 2013, AA, 556, A36.

Grossman, L., 1972, Geochim Cosmochim, Acta, 36, 597.

Grundy, W.M., & The New Horizons Science team, 2016, Nature, 539, 65.

Hartmann, L., & Stauffer, J.R., 1989, AJ, 97, 873

Herbig, G.H., 1957, ApJ, 125, 612.

Hoyle, F., 1960, Q. J. Roy. Astron. Soc., 1, 28.

Jeans, J., 1928, Astronomy and Cosmogony, (Cambridge Cup), p.389.

Johns-Krull, C.M., 2007, ApJ, 664, 975.

Johnson, T.V., & Lunine, J.I., 2005, Nature, 435,69.

Königl, A., 1991, ApJ, 370, L39.

Kuiper, G.P., 1951, in Astrophysics, ed. J. A. Hynek (New York: McGraw-Hill), p.387.

Laplace, P.S. de, 1796, Exposition du Système du Monde (Paris: Courcier).

Lewis, J.S., 1972, Earth Planet Sci. Letts., 15, 286.

Lewis, J.S., 1974, Sci., 186, 440.

Lewis, J.S., & Prinn, R.G., 1989, ApJ, 238, 357.

Lodders, K., 2008, ApJ, 674, 607.

Lodders, K., Palme, H., & Gail, H.-P., 2009, in Landolt-Börnstein, New Series VI/4B, ed. J.E. Trumpler (Berlin: Springer), Chap.4.4, p.1.

Miller, S.L., & Smythe, W.D., 1970, Sci., 170, 531.

Moore, J.M., & 33 others, 2018, GeoRL, 45, 8111.

Perry, M.E. & 10 others, 2011, Planet Space Sci., 59, 1925.

Porter, S.B., Buie, M.W. and 12 others, 2018, AJ, 156, Issue 1, article id. 20, 7pp.

Prentice, A.J.R., 1973, AA, 27,237.

Prentice, A.J.R., 1974, in In the beginning…, ed. J.P. Wild (Canberra: Aust. Acad. Sci.), p.15.

Prentice, A.J.R., 1978, Moon & Planets, 19, 341.

Prentice, A.J.R., 1990. Proc. Astron. Soc. Aust., 8, 364.

Prentice, A.J.R., 1991, Proc. Astron. Soc. Aust., 9, 321.

Prentice, A.J.R., 1993a, Proc. Astron. Soc. Aust., 10, 189.

Prentice, A.J.R., 1993b, Aust. J. Astron., 5, 111.

Prentice, A.J.R., 2001, Earth Moon & Planets, 87, 11.

Prentice, A.J.R., 2006, PASA, 23, 1.

Prentice, A.J.R., 2008, 39th LPSC, abstract #1945.

Prentice, A.J.R., 2011, AGU Fall Meeting, abstract #U23B-05.

Prentice, A.J.R., 2015, 46th LPSC, abstract #2664.

Prentice, A.J.R., 2016, DPS Meeting #48, id. 224.13.

Prentice, A.J.R., 2018, DPS Meeting #50, id.113.03.

Prentice, A.J.R., 2019, 233rd AAS Meeting, id. 467.01.

Prentice, A.J.R. & Dyt, C.P., 2003, MNRAS, 341, 644.

Prialnik, D., Sarid, G., Rosenberg, E.D., & Merk, R., 2008, Space Sci. Rev. 138, 147.

Schatzman, E., 1949, Bull. Acad. Roy. Belgique, 35, 1141.

Smith, D.E., Zuber, M.T., Phillips, R.J., Solomon, S.C. & 13 others. 2012, Sci., 336, 24.

Solomon, S.C., McNutt, R.L., Gold, R.E., & Domingue, D.L., 2007, Space Sci. Rev., 131, 3.

Stern, S.A., and the New Horizons Science Team, 2015, Sci., 350, 292.

Thomas, P.C., 2010, Icarus, 208, 395.

Trujillo, C.A., & Brown, M.E., 2003, Earth, Moon & Planets, 92, 99.

Turcotte, S., & Wimmer-Schweingruber, R.F., 2002, J. Geophysics. Res., (Space Phys.) 107, 1442.

Urey, H.C., 1955, Proc. Natl. Acad. Sci., 41, 127

Ustyugova, G.V., Koldoba, A.V., Romanova, M.M., & Lovelace, R.V.E., 2006, ApJ, 646, 304.

Weaver, H.A., & 50 others, 2016, Science, Issue 6279, id. aae0030.

Weidenschilling, S.J., 1977, Astrophys. Space Sci., 51, 153.

**Table 1** Orbital radii and thermophysical properties of the outer system of protosolar gas rings, including the protosolar rings shed at the orbits of Mercury and Ceres.

| Planet | $n$ | $t_n$ ($10^5$ yr) | $M_n$ ($M_\odot$) | $R_n$ ($R_\odot$) | $R_{n,p}$ ($R_\odot$) | $T_n$ (K) | $p_n$ (bar) |
|---|---|---|---|---|---|---|---|
| Quaoar  | 0 | 0.45 | 1.2133 | 7469  | 9281   | 26.26  | 1.265E-9[α] |
| Neptune | 1 | 0.95 | 1.2074 | 5353  | 6436   | 35.6   | 5.25E-9 |
| Uranus  | 2 | 1.44 | 1.1999 | 3438  | 4125   | 51.9   | 3.15E-8 |
| Saturn  | 3 | 1.99 | 1.1870 | 1727  | 2050   | 93.9   | 5.39E-7 |
| Jupiter | 4 | 2.31 | 1.1740 | 952.6 | 1118.2 | 157.7  | 6.52E-6 |
| Ceres   | 5 | 2.54 | 1.1584 | 513.5 | 594.9  | 271.7  | 8.91E-5 |
| Mercury | 9 | 2.89 | 1.0849 | 76.69 | 83.205 | 1638.2 | 1.811E-1 |

[α] E-9 means $10^{-9}$, etc.

**Table 2** Bulk chemical composition mass fractions $X_j$ and mean density $\rho_{\rm cond}$ of the unprocessed condensate in outer system of protosolar gas rings, including the ring shed at the orbit of Ceres.

| Planet orbit | $X_{\rm rock}$ | $X_{C(s)}$ | $X_{H_2O}$ | $X_{CO_2}$ | $X_{CH_4(\rm cl.)}$[α] | $X_{CH_4}$ | $\rho_{\rm cond}$ (g/cm$^3$) |
|---|---|---|---|---|---|---|---|
| Quaoar  | 0.5268 | 0.0163 | 0.1845 | 0.2211 | 0.0286 | 0.0227 | 1.80 |
| Neptune | 0.4918 | 0.0248 | 0.2408 | 0.2053 | 0.0373 | 0.0    | 1.76 |
| Uranus  | 0.4551 | 0.0310 | 0.2999 | 0.1676 | 0.0464 | 0.0    | 1.63 |
| Saturn  | 0.4343 | 0.0482 | 0.5175 | 0.0    | 0.0    | 0.0    | 1.44 |
| Jupiter | 0.5588 | 0.0211 | 0.4201 | 0.0    | 0.0    | 0.0    | 1.57 |
| Ceres   | 1.000  | 0.0    | 0.0    | 0.0    | 0.0    | 0.0    | 3.39 |

[α] Here cl. means clathrated species.

# APPENDIX

## 1. Elemental abundances in the Protosolar Cloud

Specification of the abundances of the elements in the Protosolar cloud (PSC) is essential in order to be able to (1) determine the chemical condensation sequence in the system of gas rings that are cast off from the equator of the PSC and (2) to compute the compositional mass fractions of the various compounds making up each planet and planetoid, including proto-Pluto and Ultima Thule.

**STEP 1**: Specifying the abundances of the rocky & metallic elements relative to the photospheric abundance of H

The vast wealth of information that has been gleaned from the study of meteorites, especially of primitive carbonaceous CI-chondrites, enables very accurate estimates to be made of the abundances of the rock and metal-like elements that existed in the PSC. I assume that the extensively examined Orgueil meteorite is a faithful custodian of the relative abundances of the mix of rocky & metallic elements and adopt the cosmochemical abundances that are given in Table 3 of Lodders *et al.* (2009). The logarithm of the cosmochemical abundance of Si is $\log_{10} N_{cosmo}(\text{Si}) = 6.000$.

Next, the abundances of each rocky & metallic element X relative to elemental H is converted to the astronomical scale, on which $\log_{10} N_{astron}(\text{H}) = 12.000$, using equation (1) of Lodders *et al.* (2009), namely

$$\log_{10} N_{astron}(X) = \log_{10} N_{cosmo}(\text{Si}) + 1.533$$

**STEP 2**: Specification of the abundances of the volatile elements C, N & O and of the noble gases Ne and Ar relative to H

The photospheric abundances of C, N, & O are taken from the self-consistent analyses of Caffau *et al.* (2010, 2009, 2010). Lodders (2008) has determined the photospheric abundance of Ar. These authors find:

$$\log_{10} N_{astron}(\text{C}) = 8.50 \pm 0.06$$
$$\log_{10} N_{astron}(\text{N}) = 7.86 \pm 0.12$$
$$\log_{10} N_{astron}(\text{O}) = 8.76 \pm 0.07$$
$$\log_{10} N_{astron}(\text{Ar}) = 6.50 \pm 0.10$$

Neon cannot be measured in the solar photoshere spectrum. Caffau *et al*. (2009) recommend the choice:

$$\log_{10} N_{astron}(\text{Ne}) = 8.02$$

**STEP 3**: Specifying the H abundance using helioseismology

The study of helioseismological data has allowed a precise determination of the mass abundance fraction of hydrogen in the solar atmosphere, namely $X_H = 0.73885 \pm 0.0024$, Basu & Anita (2004, 2008).

**STEP 4**: Solving for the He abundance

The data from STEPS 1-3 can now be combined to obtain a unique set of abundances for all elements in the photosphere. The helium mass fraction is $X_{He} = 0.24606$, giving $\log_{10} N_{astron}(\text{He}) = 10.92358$. The total heavy element mass fraction is $X_{heavy} = 0.015089$ and $X_{heavy}/X_H = 0.020422$. These values agree closely with those obtained by Caffau *et al*. (2010), namely $X_{heavy} = 0.0154$ and $X_{heavy}/X_H = 0.0211$. The slight difference arises from their different choice of meteoritic abundances.

**STEP 5**: Elemental abundances of the PROTOSOLAR CLOUD

The final step in the calculation is to deduce the elemental abundances in the PSC. I follow the procedure spelt out by Asplund *et al.* (2009) and assume that the PSC abundances are identical to those of the bulk Sun. The bulk Sun abundances of He and the heavy elements differ from the photospheric because of the effects of thermal diffusion and gravitational settling in the photosphere (Turcotte & Wimmer-Schweingruber, 2002. The connection between the PSC and astronomical abundances is given by the equations:

$$\log_{10} N_{PSC}(H) = \log_{10} N_{astron}(H)$$
$$\log_{10} N_{PSC}(He) = \log_{10} N_{astron}(He) + 0.05$$
$$\log_{10} N_{PSC}(X) = \log_{10} N_{astron}(X) + 0.04$$

Here X refers to any element heavier than He. Atomic weights have been taken from Eidelman *et al.* (2004).

Table 1: Elemental abundances in the Protosolar cloud

| Atomic number | Symbol X | Abundance $\log_{10} N_{PSC}(X)$ | Atomic weight |
|---|---|---|---|
| 1 | H | 12.0000000 | 1.00794 |
| 2 | He | 10.9735794 | 4.0026002 |
| 3 | Li | 3.31808 | 6.941 |
| 5 | B | 2.84716 | 10.811 |
| 6 | C | 8.54000 | 12.0107 |
| 7 | N | 7.90000 | 14.00674 |
| 8 | O | 8.80000 | 15.9994 |
| 9 | F | 4.47826 | 18.99840 |
| 10 | Ne | 8.06000 | 20.1797 |
| 11 | Na | 6.32887 | 22.98977 |
| 12 | Mg | 7.58584 | 24.3050 |
| 13 | Al | 6.49051 | 26.98154 |
| 14 | Si | 7.57300 | 28.08550 |
| 15 | P | 5.48628 | 30.97376 |
| 16 | S | 7.21447 | 32.066 |
| 17 | Cl | 5.28649 | 35.4527 |
| 18 | Ar | 6.54000 | 39.948 |
| 19 | K | 5.13529 | 39.0983 |
| 20 | Ca | 6.35404 | 40.078 |
| 21 | Sc | 3.10956 | 44.955910 |
| 22 | Ti | 4.96570 | 47.867 |
| 23 | V | 4.02016 | 50.9415 |
| 24 | Cr | 5.70010 | 51.9961 |
| 25 | Mn | 5.53773 | 54.93805 |
| 26 | Fe | 7.51252 | 55.845 |
| 27 | Co | 4.92518 | 58.93320 |
| 28 | Ni | 6.25695 | 58.6934 |
| 29 | Cu | 4.30620 | 63.546 |
| 30 | Zn | 4.68694 | 65.39 |
| 31 | Ga | 3.13648 | 67.723 |
| 32 | Ge | 3.64488 | 72.61 |

| 33 | As | 2.35833 | 74.9216 |
|---|---|---|---|
| 34 | Se | 3.40223 | 78.96 |
| 35 | Br | 2.60238 | 79.904 |
| 36 | Kr | 3.32000 | 83.80 |
| 37 | Rb | 2.42426 | 85.4678 |
| 38 | Sr | 2.94222 | 87.62 |
| 39 | Y | 2.22814 | 88.90585 |
| 40 | Zr | 2.59003 | 91.224 |
| 42 | Mo | 1.99788 | 95.94 |
| 52 | Te | 2.24417 | 127.6 |
| 54 | Xe | 2.31 | 131.29 |
| 56 | Ba | 2.23670 | 137.327 |
| 82 | Pb | 2.09544 | 207.2 |
| 90 | Th | 0.1183071 | 232.0381 |
| 92 | U | -0.4761485 | 238.0289 |

<u>Note</u>: The above table excludes all elements X for which $\log_{10} N_{PSC}(X) < 2.0$. That is, we ignore elements for which $N_{PSC}(X)/ N_{PSC}(H) < 10^{-10}$. Mo is included in the tabulation as its sits at the lower limit $N_{PSC}(X) = 2.0$.

**STEP 6**: Mass fractions $X_i$ of the major chemical constituents of proto-Pluto

The table below gives the mass fraction of the principal chemical constituents that potentially exist in the protosolar cloud, based on groups according as sulphides, halides, oxides, heavy inert elements or as significant unattached elements. H and He are listed at the top of the Table. The sulphides, oxides and unattached elements are further divided into 2 sub-groups of major and minor. Major elements have PSC abundance $N_{PSC}(X) > 10^4$ and minor elements have $10^2 < N_{PSC}(X) < 10^4$. With the 2 exceptions of U & Th, all remaining elements (each having PSC abundance $N_{PSC}(X) < 10^2$) have been placed in a residual group at the foot of the Table. The notation 1E-08 means $1 \times 10^{-8}$, etc.

**Table 2. Mass fractions of chemical constituents of protosolar cloud material**

| Chemical constituent | Mass fraction |
|---|---|
| Hydrogen | **0.714444393** |
| Helium | **0.267659510** |
| <u>Heavy inert gases</u>:  Neon  Argon  Kr, Xe  Sub-total: | 1.6465525E-03  9.843644E-05  1.4347E-07  **1.745132E-03** |
| <u>Major sulphides</u>:  MnS  CoS  ZnS  FeS  NiS  Sub-total: | 2.132707E-05  5.44353E-06  3.36832E-06  8.8093455E-04  1.1654659E-04  **1.027620E-03** |
| <u>Minor sulphides</u>:  AsS  MoS$_2$ | 1.735E-08  1.132E-08 |
| SnS | 1.443E-08 |
| PbS | 2.118E-08 |
| Sub-total: | **6.4E-08** |
| Halides: | |
| NaCl | 8.03304E-06 |
| NaBr | 2.927E-08 |
| $CaF_2$ | 8.3445E-07 |
| Sub-total: | **8.897E-06** |
| Major oxides: | |
| $Na_2O$ | 4.269398E-05 |
| MgO | 1.1036987E-03 |
| $Al_2O_3$ | 1.1209123E-04 |
| $SiO_2$ | 1.5974301E-03 |
| $P_2O_5$ | 1.545373E-05 |
| $K_2O$ | 4.57042E-06 |
| CaO | 8.945106E-05 |
| $TiO_2$ | 5.24468E-06 |
| $V_2O_5$ | 6.7698E-07 |
| $Cr_2O_3$ | 2.707396E-05 |
| $Fe_3O_4$ | 1.0117579E-03 |
| Sub-total: | **4.010143E-03** |
| Minor oxides: | |
| $Li_2O$ | 2.209E-08 |
| $Sc_2O_3$ | 6.306E-08 |
| $Ga_2O_3$ | 9.120E-08 |
| $GeO_2$ | 3.2818E-07 |
| SrO | 6.446E-08 |
| $ZrO_2$ | 3.407E-08 |
| BaO | 1.879E-08 |
| Sub-total: | **6.22E-07** |
| Unattached elements: | |
| Major element: Cu | 9.1400E-07 |
| Minor elements: | |
| B | 5.40E-09 |
| Se | 1.4170E-07 |
| Rb | 1.613E-08 |
| Y | 1.068E-08 |
| Te | 1.591E-08 |
| Unattached total: | **1.104E-06** |
| Icy Hydrides: | |
| $CH_4$ | 3.9530457E-03 |
| $NH_3$ | 0.9613683E-03 |
| $H_2O$ | 6.1880391E-03 |
| Sub-total: | **1.11024531E-02** |
| Residual elements: | **6.2E-08** |
| **GRAND TOTAL:** | **1.000000000** |

The protosolar mass fractions of the long-lived radioactive elements Th & U are shown below. Their mass fractions are included in the total mass of residual elements that is shown at the end of Table 2.

Table 3. Mass fractions of radioactive elements in the protosolar cloud

| Element | Mass fraction |
|---------|---------------|
| Th | 2.165346E-10 |
| U | 5.651218E-10 |

## 2. Determination of the adiabatic polytropic index of the undissociated protosolar gas

Procedure. All H of the undissociated outer layers of the protosolar cloud (PSC) is present as $H_2$, all C is assumed to be tied up as CO (not $CH_4$), and all N exists as $N_2$ (not $NH_3$). Oxygen that is not tied up in rock oxides or in CO is present as $H_2O$. Let us now compute the adiabatic polytropic index of this gas mixture. The Table below shows the mass fractions $X_i$ of the principal molecular species $H_2$, $N_2$, $H_2O$ and CO, along with the mass fractions of the 3 main inert gases He, Ne and Ar. Here $i$ denotes the species index. The mass fractions $X_i$ {$i$ = 1,7} are determined from Table 2. Column 4 of the table gives the molecular weight $\mu_i$ of each species. Column 5 gives the number of degrees of atomic freedom $\nu_i$ of each of the molecular or inert gas species.

| Species index $i$ | Molecule/ inert gas atom | Mass fraction $X_i$ | Molecular weight $\mu_i$ | Degrees of atomic freedom $\nu_i$ |
|---|---|---|---|---|
| 1 | $H_2$ | 0.7161055 | 2.01588 | 5 |
| 2 | $N_2$ | 0.0007907 | 28.01348 | 5 |
| 3 | CO | 0.0069020 | 28.01010 | 5 |
| 4 | $H_2O$ | 0.0017488 | 18.01528 | 6 |
| 5 | He | 0.2676595 | 4.0026002 | 3 |
| 6 | Ne | 0.0016466 | 20.1797 | 3 |
| 7 | Ar | 0.0000984 | 39.948 | 3 |

The average number of degrees of atomic freedom of the mixture $\langle \nu \rangle$ is given by

$$\langle \nu \rangle = \left\{ \sum_{i=1}^{7} \frac{\nu_i X_i}{\mu_i} \right\} \bigg/ \left\{ \sum_{i=1}^{7} \frac{X_i}{\mu_i} \right\} = 4.683$$

The adiabatic polytropic index of the undissociated gas is thus

$$n_{ad} = \tfrac{1}{2} \langle \nu \rangle = 2.342$$

The mean molecular weight of the above gas mixture is

$$\langle \mu \rangle = \left\{ \sum_{i=1}^{7} X_i \right\} \bigg/ \left\{ \sum_{i=1}^{7} \frac{X_i}{\mu_i} \right\} = 2.355$$

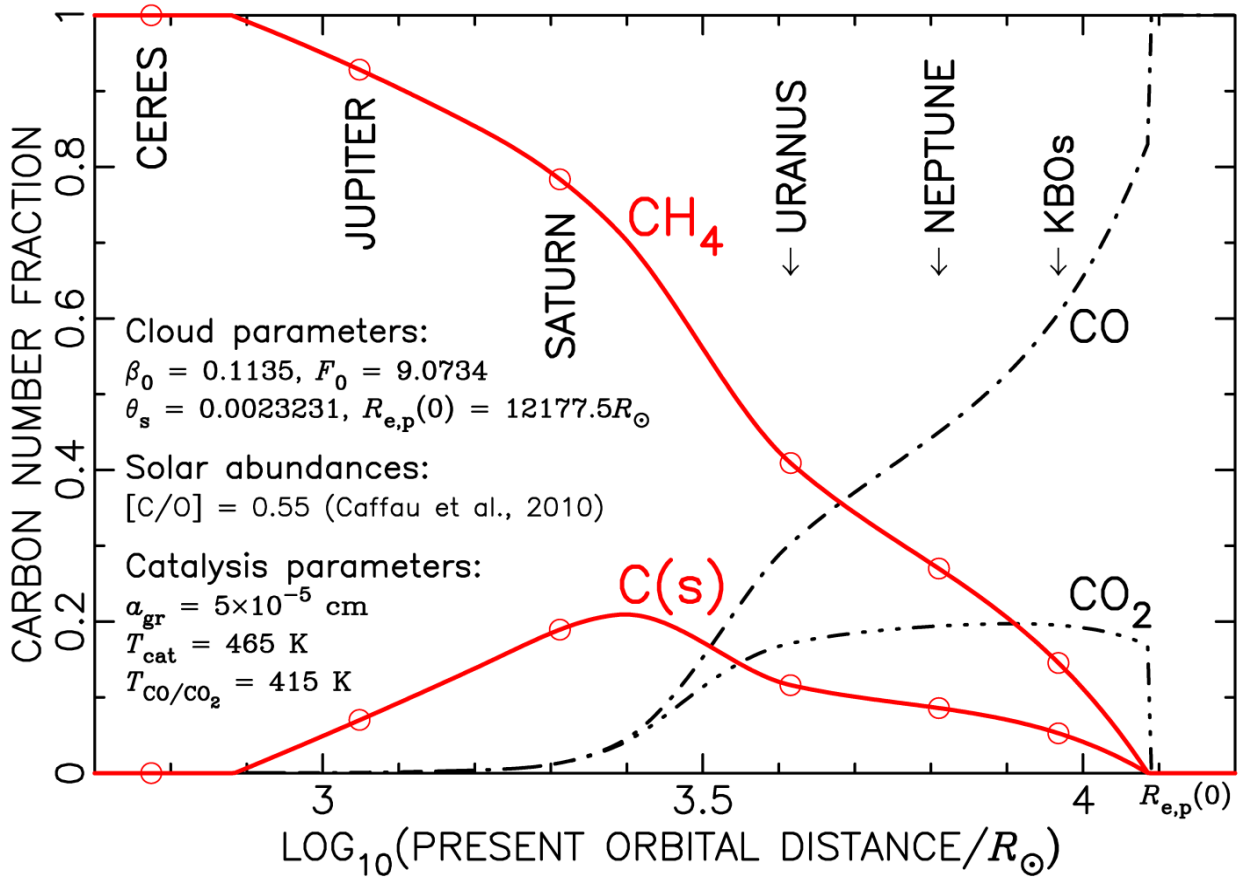

**Figure 1**

This graph shows the distribution of carbon number between the 4 possible states of CO, $CO_2$, $CH_4$ and C(s). The carbon numbers have been computed for the gas at the equator of the contracting Protosolar Cloud (PSC) and plotted against the cloud present-day equivalent equatorial radius $R_{e,p}$. Initially all carbon is present as CO. A steady synthesis of $CH_4$ and graphite C(s) from CO and $CO_2$ takes place as a result of catalysed reactions on the surfaces of pure metallic grains consisting of Fe, Ni and Co. As explained in Sections 4.1 and 5.1, the present orbital distances differ from the initial orbital values $R_e$ because of secular expansion. This is due to loss of the mass $M_e$ of the PSC during its gravitational contraction to present solar size. We have $R_{e,p} = R_e \cdot (M_e/M_\odot)$. The initial cloud mass is $M_e(0) = 1.21775 M_\odot$.

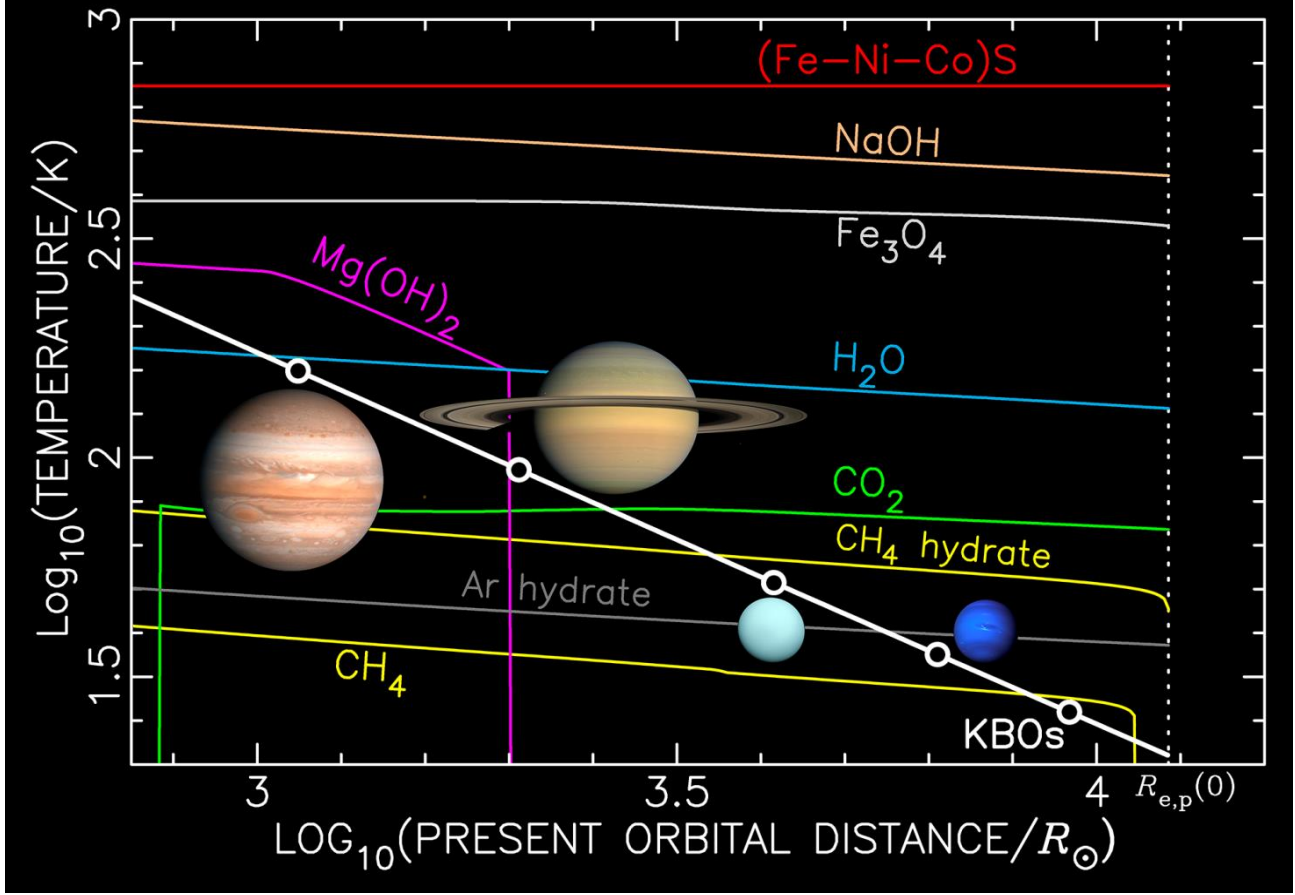

**Figure 2**

The heavy white curve in this diagram shows the temperature at the equator of the Protosolar Cloud (PSC), plotted against the present-day equivalent equatorial radius $R_{e,p}$. The other fainter curves show the condensation temperatures of the principal ices $H_2O$, $CO_2$ and $CH_4$, as well as those of the low temperature rock phases. These temperatures have been computed for the gas pressure $p_n(0)$ on the mean central orbit of a locally detached gas ring having mean orbital radius equal to the equatorial radius $R_e$ of the PSC.

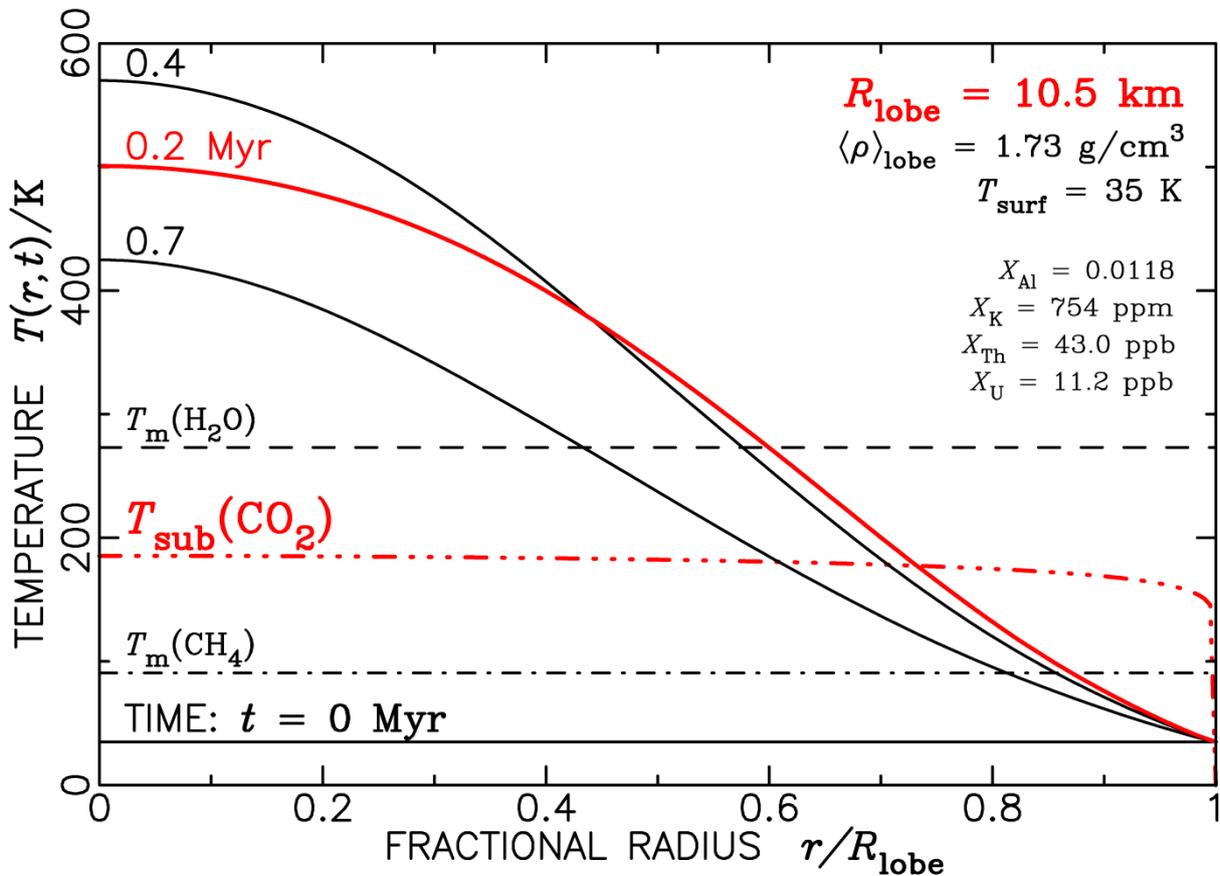

**Figure 3**

This graph shows the run of temperature $T(r,t)$ versus radius $r$ inside a spherical lobe of Ultima Thule having radius 10.5 km. The temperature profiles are plotted at several key times $t$ during the thermal evolution, starting from time $t = 0$ where the internal temperature is everywhere set equal to 35 K. The heating is due to the decay of short-lived radioactive isotope $^{26}$Al which has a half-life of 0.717 Myr. The melting curves for $H_2O$ and $CH_4$ ice are shown, along with the sublimation temperature of $CO_2$ ice.

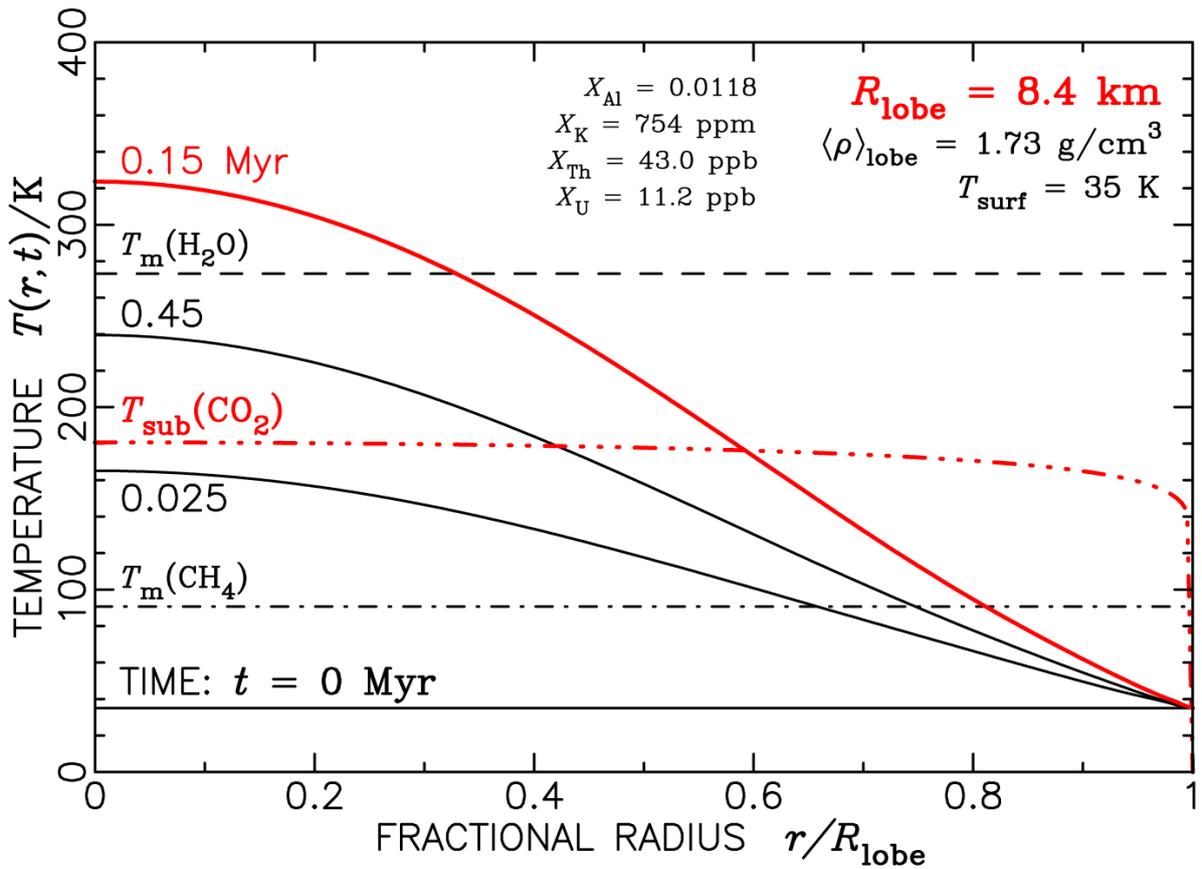

**Figure 4**

This graph shows plots of the temperature profiles $T(r,t)$ versus radius $r$ inside a smaller 8.4 km radius lobe of Ultima Thule, at several key times $t$. Again all heating is due to the decay of short-lived radioactive isotope $^{26}$Al which has a half-life of 0.717 Myr. The melting curves for $H_2O$ and $CH_4$ ice are shown, along with the sublimation temperature of $CO_2$ ice. We note the temperatures within this smaller lobe do not rise to same peak values as those within the larger lobe of radius 10.5 km, shown in Figure 3.

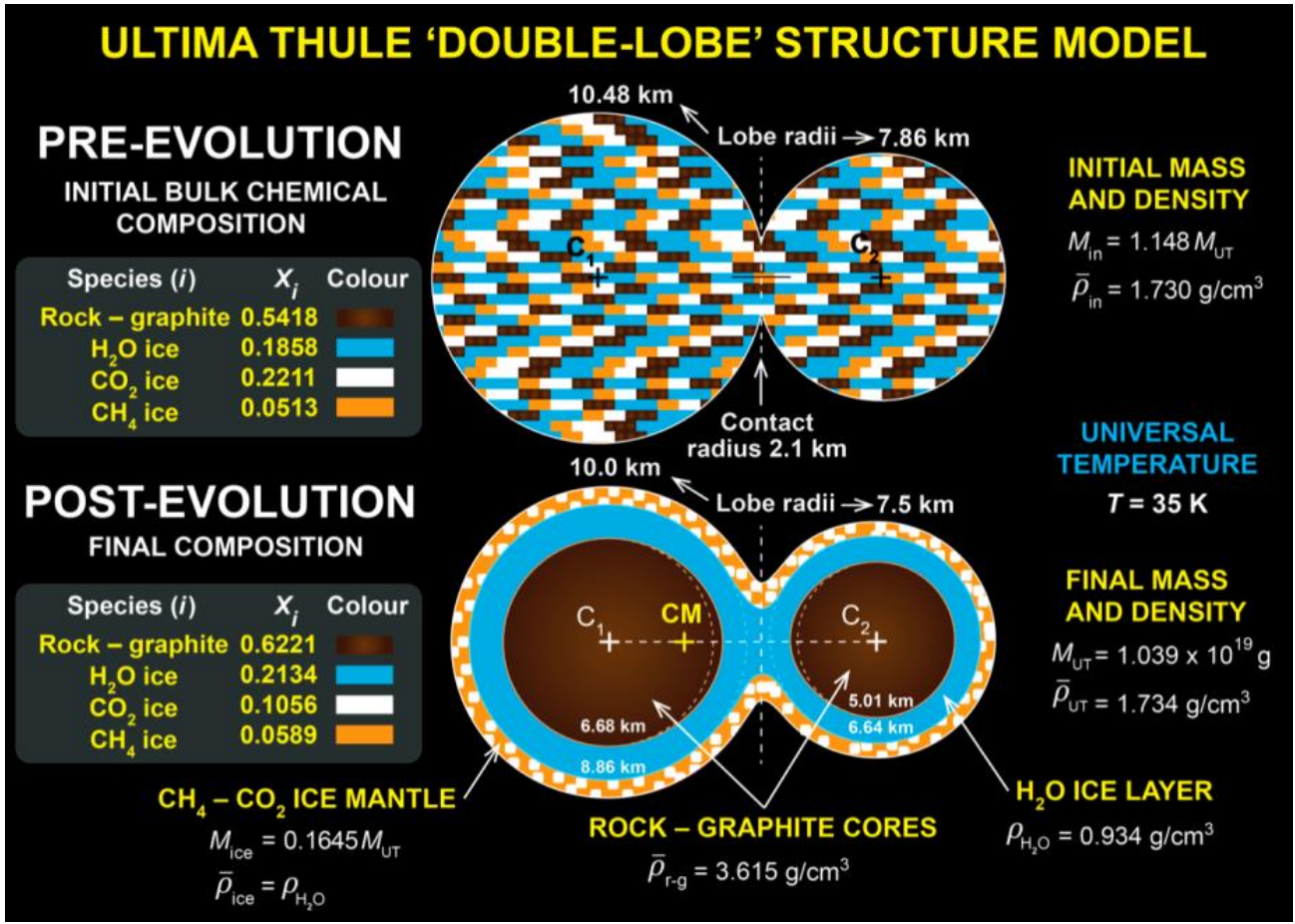

**Figure 5**

This figure is a schematic illustration of the internal physical structure of Ultima Thule, assuming that consists of two contacting lobes having present day radii of 10 km and 7.5 km. Prior to the onset of radiogenic heating due to $^{27}$Al decay , each lobe contained large stores of $CH_4$ and $CO_2$ ice. These made up 5% and 22% of their masses, respectively. Radiogenic heating first causes the $CH_4$ ice to melt. The liquid $CH_4$ migrates quickly to the surface to form a thick crystalline shell around each lobe. It forms a smooth waist profile at the contact surface. Subsequently, sublimation of $CO_2$ ice takes place. The $CO_2$ vapours rise to the surface where they become trapped by the outer crystalline $CH_4$ ice shell. This outer shell is explosively disrupted by the build-up of gas pressure, allowing much of the $CO_2$ to escape. In this schematic it is assumed that 60% of the total store of each lobe's $CO_2$ escapes into space. The remaining 40% is trapped as ice in the outer $CH_4$ ice shell.